\documentstyle[aps,prl]{revtex}
\begin{document}
\draft
\twocolumn[\hsize\textwidth\columnwidth
\hsize
%Phys. Rev. B MS #  BA7138
%submitted electronically January 6, 1999
\csname @twocolumnfalse\endcsname

\title{Order Parameter Symmetries in High
Temperature Superconductors}
\author{R. A. Klemm,$^1$ C. T. Rieck$^2$ and K.
Scharnberg$^2$}
\address{$^1$Materials Science Division, Argonne
National Laboratory, Argonne, IL
60439 USA}
\address{$^2$Fachbereich Physik, Universit{\"a}t
Hamburg, Jungiusstra\ss e 11, D-20355 Hamburg,
Germany}
\date{\today}
\maketitle
\begin{abstract}
The symmetry operations of the crystal groups
relevant for
the high temperature superconductors
HgBa$_2$CuO$_{4+\delta}$ (Hg1201),
YBa$_2$Cu$_3$O$_{7-\delta}$ (YBCO),
and Bi$_2$Sr$_2$CaCu$_2$O$_{8+\delta}$ (BSCCO) are
elucidated.
The allowable combinations of the superconducting
order
parameter (OP) components are presented for both the angular momentum
and lattice representations.  For tetragonal Hg1201, the
spin singlet OP components are composed from four sets of compatible
basis functions, which combine to give the generalized $s$-,
$d_{x^2-y^2}$-, $d_{xy}$-, and $g_{xy(x^2-y^2)}$-wave OPs.  Although
YBCO and BSCCO are both orthorhombic, they allow different mixing of these
components.    In YBCO, the elements of those $s$- and
$d_{x^2-y^2}$-wave sets (and of those $d_{xy}$- and
$g_{xy(x^2-y^2)}$-wave sets) are compatible, but in BSCCO, elements of
$s$- and $d_{xy}$- wave sets (and of $d_{x^2-y^2}$- and
$g_{xy(x^2-y^2)}$-wave sets) are compatible.
The
Josephson critical current density $J_c^J$ across $c$-axis twist
junctions
 in the
vicinity of $T_c$ is then evaluated as a function of the
 twist
angle $\phi_0$, for each allowable OP combination, for both coherent
and incoherent tunneling.  For tight-binding Fermi surfaces, we argue
that coherent tunneling is only possible for $\phi_0=0^{\circ},
90^{\circ}$, for which the rotated Fermi surfaces line up.
Recent experiments of Li {\it et al.} demonstrated the independence of
$J_c^J(\phi_0)/J_c^S$ upon $\phi_0$ at and below $T_c$, where $J_c^S$ is the critical
current density of a constituent single crystal.
These experiments are shown
to be consistent with an OP containing an $s$-wave
component, but {\it inconsistent} with an OP containing the
purported $d_{x^2-y^2}$-wave component.  In addition, we argue that
they demonstrate that the interlayer tunneling across untwisted
layers in single crystal BSCCO is entirely {\it incoherent}, with no
 amount of forward scattering.	We propose a new type of
tricrystal experiment using single crystal $c$-axis twist junctions, that does
not employ substrate grain boundaries.
\end{abstract}
\vskip0pt
\vskip0pt\vskip0pt
]
\narrowtext

\section{Introduction}

There have been some exciting new developments
which reopen the question of the symmetry of the order parameter (OP) in
the high temperature superconducting compounds (HTSC).
\cite{Li1,Li2,Li3,Kleiner,Goldman,Woods} Although previous experiments such as
the penetration depth, \cite{Bonn}
 and ``phase-sensitive'' tricrystal experiments were
interpreted
as giving strong evidence for an
OP in YBa$_2$Cu$_3$O$_{7-\delta}$ (YBCO) that was
dominated
at low temperature $T$ by a component of
$d_{x^2-y^2}$ symmetry, \cite{Kirtley} other experiments
showed that there was also a substantial $s$-wave
component
 in that material as well, \cite{Dynes,Kousnetzov} and a very recent
experiment suggested a nodeless OP. \cite{Goldman}
However, as discussed in the following, the crystal
symmetry of YBCO is orthorhombic, with distinct
Cu-O bond lengths $a$ and $b$
within the CuO$_2$ planes,  which are normal to and along
the CuO chain direction, respectively.	This particular
type of orthorhombicity then allows for an arbitrary mixing of
$d_{x^2-y^2}$- and
$s$-wave OP components, even at $T_c$. \cite{Kousnetzov,Goldman}

To date, the only hole-doped cuprate superconducting material known to
have a consistently tetragonal crystal structure is
HgBa$_2$CuO$_{4+\delta}$ (Hg1201).  Unfortunately,
 no ``phase-sensitive'' experiments  have yet been
performed on this material.  The only experiment relevant to
the symmetry of the OP was point contact
tunneling, which appeared consistent with an $s$-wave OP at
low $T$. \cite{Chen}  A material which can
sometimes be made in a tetragonal structure is
Tl$_2$Ba$_2$CuO$_{4+\delta}$ (Tl2201).	Point contact
tunneling and Raman scattering on Tl2201 appeared to
be consistent with line nodes of the OP, as
expected for a $d_{x^2-y^2}$-wave OP. \cite{Zasadzinski,Kang}
 There was one
phase-sensitive experiment performed on Tl2201, which
 claimed to give strong evidence for
a $d_{x^2-y^2}$-wave OP in that material at
low $T$. \cite{Tsuei}. However,  Tl2201 is
difficult to prepare in the tetragonal form, and most samples
of it are actually orthorhombic, in a form similar to that of YBCO.
 \cite{Jorgensen2} The electron-doped cuprate Nd$_{2-x}$Ce$_x$CuO$_{4-\delta}$
(NCCO) is also tetragonal, and microwave experiments suggested a full
BCS-like gap, \cite{Wu} suggestive of conventional $s$-wave
superconductivity in that material.  However, new Pb/NCCO $c$-axis Josephson
tunneling experiments demonstrate a surprising small $s$-wave OP
component, and no evidence of a gap, suggesting that the full OP might
not be pure $s$-wave. \cite{Woods}  

With regards to Bi$_2$Sr$_2$CaCu$_2$O$_{8+\delta}$, (BSCCO),
recent phase-sensitive	tricrystal experiments also provided evidence at
the tricrystal intersection point for a half-integral
flux quantum $\Phi_0/2$, where the standard superconducting
quantum of flux is $\Phi_0=hc/2e$. \cite{Kirtley2} In addition,
angle-resolved photoemission spectroscopy (ARPES), penetration depth
measurements, point contact tunneling measurements, and
Raman scattering experiments were all interpreted as providing additional
 evidence for a superconducting OP consistent with $d_{x^2-y^2}$
symmetry, although such non-phase-sensitive
experiments could  only detect the magnitude of the OP.
\cite{Ding,Shen,Sridhar,Miyakawa,Devereaux}

Early attempts to observe $c$-axis Josephson tunneling
between
Pb and BSCCO were unsuccessful. \cite{Beasley}. Those experimenters
prepared
 thin films of BSCCO, and deposited a thick
layer (roughly 1000{\AA}) of Ag on top of it,
followed
by a thicker layer of Pb.  However, new $c$-axis Josephson
tunneling between Pb and
BSCCO provided compelling evidence for	a small $s$-wave
OP component
 at low $T$. \cite{Kleiner} These experimenters cleaved a single
crystal
of BSCCO, and deposited only about 10{\AA} of
Ag on it prior to depositing Pb.  The product of
 the
$c$-axis superconducting critical current $I_c$ and the
 quasiparticle resistance $R_n$ was
found to be very small, about 1-2 $\mu$V, curiously about the same as
for the new Pb/NCCO junctions. \cite{Woods}  Nevertheless,
this evidence for
(at least) a small $s$-wave OP component in BSCCO at low
$T$ is very interesting, since
 BSCCO is
orthorhombic in a different sense than is YBCO.
 As discussed in the following,
this form of orthorhombicity
does {\it not} allow for a mixing of $d_{x^2-y^2}$- and
$s$-wave symmetry at $T_c$.  Thus, these $c$-axis
Pb/BSCCO
Josephson experiments could only be explained in the
``$d$-wave scenario''
(in which the purported $d_{x^2-y^2}$-wave OP component
were dominant at $T_c$) by the appearance of a second
phase transition in BSCCO between the $T_c$ values
of BSCCO and Pb.  The same argument would of course apply to the
$d$-wave scenario as an explanation of the new Pb/NCCO experiments. \cite{Woods}

Furthermore,   $c$-axis twist experiments have
been performed, in
which a single crystal of BSCCO was cleaved in the
$ab$-plane,
the two
single crystal pieces twisted an angle $\phi_0$ with respect
to
each other, and fused together just below the melting
temperature. \cite{Li1,Li2,Li3}
Preliminary experiments performed at low $T$ and large magnetic
fields ${\bf H}||\hat{\bf c}$ showed that the measured
$I_c$ across the twist junction did not depend upon
 $\phi_0$. \cite{Li1}  However, these early
results could be interpreted in terms of a dominant
$d_{x^2-y^2}$-wave OP component that
could ``twist'' by mixing in a subdominant $d_{xy}$-wave OP
component. \cite{KRS}	In the vicinity of a 45$^{\circ}$
twist junction, the  dominant $d_{x^2-y^2}$-wave OP component
would be suppressed, allowing the subdominant
$d_{xy}$-wave
OP component to become non-vanishing well above $T_{c2}$,
even extending to some extent up to $T_c$.  Thus, one could
have a
rather $\phi_0$-independent $I_c(\phi_0)$ behavior
 in
the vicinity of the measurement $T=10$ K,
as observed. \cite{Li1}   However, such a scenario would
necessarily
{\it also} imply a second bulk phase transition at $T_{c2}<<T_c$,
below which  the $d_{xy}$-wave
 form could appear. $T_{c2}$
would have to be very low ({\it i. e.}, $T_{c2}<1$K)
in
the bulk of the sample,
if the states in the superconducting ``gap'' evident at low
$T$
were correctly interpreted as  arising from  nodes in the
purported $d_{x^2-y^2}$-wave OP.
\cite{Ding,Shen,Sridhar,Miyakawa,Devereaux}  In addition,
although such a scenario could lead to a non-vanishing
 $I_c(\phi_0=45^{\circ})$ in the vicinity of $T_c$, it could
not
lead to an isotropic $I_c(\phi_0)$ near to $T_c$.

Now the $c$-axis twist experiments have been
performed just below $T_c$ in BSCCO. \cite{Li2,Li3}  These
new experiments
{\it also} show no $\phi_0$ dependence of $I_c(\phi_0)$.
In these experiments, $I_c$
was found to scale with the twist junction area, indicating uniform
junctions, and the critical current density $J_c^J$ of the twist
junctions was found to be the same as $J_c^S$, the critical current
density
for the constituent single
crystals, regardless of $\phi_0$.  Since BSCCO is intrinsically a
stack of Josephson junctions, \cite{Mueller} these experiments
demonstrate that the twist junctions behave precisely as the
Josephson junctions intrinsic to single crystal BSCCO.
Based upon our earlier detailed calculations, these
experiments
are incompatible with the ``$d$-wave scenario.'' \cite{KRS}
Here, we present general arguments, based solely  upon
the
crystal symmetry, which demonstrate the basic
incompatibility
of a
$d_{x^2-y^2}$-wave with an $s$-wave  OP component.
As shown
in the following, the
 experiments of Li {\it et al.} provide strong evidence
that
 the OP in BSCCO contains the $s$-wave component at
$T_c$.	We employ group theory to
 argue that this experiment demonstrates that the
dominant OP
at $T_c$ contains the $s$-wave component, but not the
purported $d_{x^2-y^2}$-wave component.
These new phase-sensitive experiments are thus in direct contradiction
with the phase-sensitive tricrystal experiments of Tsuei and
Kirtley {\it et al.},  which were claimed to
provide strong evidence for a dominant $d_{x^2-y^2}$-wave
OP in BSCCO at low $T$. \cite{Kirtley2}

\section{Crystal Symmetries}
Most workers in the field of HTSC agree that the superconducting
properties of the materials are determined by the
structure of the CuO$_2$ planes.  This is because the
electronic structure calculations indicate that the CuO$_2$
plane bands cross the Fermi energy $E_F$, \cite{Freeman} and
ARPES
measurements well above $T_c$ provide strong support
for these results. \cite{Olson} Within the CuO$_2$ planes,
the Cu and O
ions are situated approximately as pictured in Fig. 1a.\cite{Jorgensen}
The Cu ions are located  on a square
lattice, with O ions between near-neighbor Cu ions.

If one  ignores the complications of $c$-axis variations
in the relative
orientations of different CuO$_2$ planes,  the crystal point
group
describing a single,  tetragonal, CuO$_2$ plane is
$C_{4v}$. \cite{Tinkham,Falicov,Jorgensen3,AGL,AGR}
The $C_{4v}$ group symmetry
operations are indicated in Fig. 1b.  In addition to the
	identity operation $E$, there
are mirror reflections $\sigma_x$ and $\sigma_y$ in the
planes normal to the layers and containing
the $x$
and $y$ axes, respectively, mirror reflections $\sigma_{d1}$
and $\sigma_{d2}$ in the planes normal to the layers and
containing
the diagonals $d_1$ and $d_2$, respectively, and rotations
$C_4$,
$C_4^{-1}$, and $C_2=C_4^2$ by 90$^{\circ}$, -90$^{\circ}$,
 and 180$^{\circ}$ about the
$c$-axis, respectively.   Tinkham used the
slightly different notation $\sigma_v$, $\sigma_v'$,
$\sigma_d$, and $\sigma_{d}'$, for the four mirror
reflections, respectively. \cite{Tinkham}  This crystal symmetry is
appropriate for Hg1201, NCCO, and
for tetragonal samples
of Tl2201. \cite{AGL}

In YBCO, there is an orthorhombic distortion involving
the
CuO chains, which causes the  CuO$_2$ crystal
structure to be longer along the $b$- (or $y$-) axis than
along
the
$a$- (or $x$-) axis.  The crystal space group is
$Pmmm$, with the point group of the CuO$_2$ planes being $C^1_{2v}$, an
example of $C_{2v}$.
  For this lower symmetry, the
allowable group operations in addition to $E$ are only
$\sigma_x$, $\sigma_y$, and $C_2$, as sketched in Fig. 1c.  This
group is
also appropriate for the double chain compound
YBa$_2$Cu$_4$O$_8$ (Y124).  However, there are some
interesting distortions, which complicate this structure.
 These are the buckling of the CuO$_2$ planes along the
$c$-axis direction. \cite{Jorgensen3}  Nevertheless, there
isn't any buckling within
the CuO$_2$ planes, so it doesn't affect the symmetry arguments relevant
to the
CuO$_2$ planes.

In order to correctly interpret the new experiments on BSCCO, it is
imperative that the crystal symmetry be understood as thoroughly as
possible.  The discussion of the symmetry of the superconducting OP in
this paper depends crucially upon the notion that there is at least
one mirror plane that intersects the CuO$_2$ plane.  This mirror plane
then distinguishes OP basis functions that are odd or even with
respect to reflections in it.  As discussed in the following, the
$bc$-plane
is a strict crystallographic mirror plane in most samples.

For overdoped or optimally doped
BSCCO,
the  crystal structure is  also orthorhombic, but in this case
 the inequivalent $a$- and $b$-axes are not along the
Cu-O bond directions, but along the directions ${\bf d_1}$
and ${\bf d_2}$
{\it diagonal} to them.  In addition, there is
a periodic lattice distortion ${\bf Q}$ along one of these
diagonals, which we  identify as the $b$-
(or $d_2$-) axis,
consistent with the identification of most workers,
\cite{Moss1,Moss2}  although some workers
 identified ${\bf Q}$ as being along the $a$- (or
$d_1$-) axis direction. \cite{Coppens1,Coppens2}   More precisely,
${\bf Q}$ is incommensurate along the reciprocal lattice
wavevector
${\bf b}^*$, but commensurate in the other crystal directions,
${\bf Q}=(0,0.212,1)$. \cite{Moss1,Moss2,Coppens1,Coppens2}
  The
fact that there is a non-vanishing commensurate
$c$-axis
 component actually removes the  $\sigma_{d1}$ mirror
plane crystal symmetry, consistent with the appearance
of weak
(100)
reflections in the x-ray diffraction data. \cite{Moss2,Coppens1}.
 In addition, there is some dispute as to the precise details of
the periodic lattice distortion.  An early electron diffraction
study of BSCCO indicated that in one sample, ${\bf Q}=(0.004,0.212,1)$,
 containing
a small ${\bf a}^*$ component  that
would also remove the $\sigma_{d2}$ mirror plane symmetry, \cite{Moss1}
  although this slight ${\bf a}^*$ component
 of  ${\bf Q}$	was not seen in high
resolution x-ray diffraction experiments on related
samples. \cite{Moss1}
  More to the point,
 electron diffraction studies of the samples used in the
present $c$-axis twist experiments demonstrated that
the small ${\bf a}^*$ component of ${\bf Q}$
 was absent in 90\% of the samples. \cite{Zhu}
Thus, it appears that although the crystal  symmetry of
BSCCO is lower than orthorhombic, as the mirror plane
$\sigma_{d1}$ is not precise, most presently-made BSCCO
 samples  exhibit the $\sigma_{d2}$ mirror plane
(the $bc$-plane). Further experiments to clarify this situation
are under way. \cite{Zhu}

This periodic lattice distortion also gives rise to some
buckling of the lattice, including Cu-O bond buckling
 in the
$bc$-plane.	  Both x-ray studies of the crystal structure of BSCCO
used a four-dimensional analysis procedure, and listed the basic space group
as
$Bb2b$ (or equivalently, $A2aa$), which is the orthorhombic
point group  $C_{2v}^{13}$, another form of $C_{2v}$.
 \cite{Moss1,Moss2,Coppens1,Coppens2}	Unfortunately,
they
differ slightly in listing the full crystal group
 as $M:A2aa:\overline{1}11$ and $N_{111}^{Bb2b}$, respectively.
  In addition, there is a
further complication in some heavily underdoped samples,
in that they appear to undergo a phase transition from
orthorhombic to monoclinic. \cite{Gough1,Gough2}
 In such cases, there are no mirror
planes (other than parallel to the $ab$-plane) remaining,
 and the only non-trivial group operation
appropriate for a single CuO$_2$ plane would be
$C_2$.	However, since the $c$-axis twist experiments did
not
involve such monoclinic single crystals, \cite{Li1,Zhu} we
shall
not discuss such very low symmetry cases further.

Neglecting  these fine details,  the allowed
symmetry operations of	the simplified point group $C_{2v}^{13}$
are $E$, $\sigma_{d1}$, $\sigma_{d2}$, and $C_2$, as
illustrated in Fig. 1d. This symmetry is also appropriate for
 the compound La$_{2-x}$Sr$_x$CuO$_{4+\delta}$ (LSCO).
With regard to a single CuO$_2$ layer, the $\sigma_{d1}$ mirror-plane symmetry is strict, as
it is only broken by the $c$-axis component of ${\bf Q}$ in BSCCO.
More important,
 $C_{2v}^1$ (for YBCO) and $C_{2v}^{13}$ (for BSCCO)
differ only in
that the in-plane crystal axes are effectively rotated
by 45$^{\circ}$ about the $c$-axis from each other. As
 regards the orbital symmetry
of the superconducting OP, this difference is {\it crucial}, as noted in detail below. 

\section{Order Parameter Components}
The role of symmetry upon the OP
components in the HTSC was discussed in a recent review article by Annett,
Goldenfeld, and Leggett (AGL). \cite{AGL}
However, the notation  used by AGL was
rather confusing, such as using ``$s^{-}$'' to denote a state that
contains no $s$-wave component, and whose
leading term is $g$-wave.  More important, the question of whether
different OP components could mix with or without a second phase
transition was stated incorrectly in the conclusion.
 Therefore, as the symmetry of the OP in the HTSC is still very
much under debate, it is imperative to provide a clear discussion of
the issues.

Quantum mechanics, Bloch's theorem, and group theory dictate that the
superconducting OP must be consistent with
the crystal structure. \cite{Tinkham,AGL}  Assuming that
 the CuO$_2$ planes are
responsible for the superconductivity in the HTSC,
then the point group structure of the CuO$_2$
planes will identify those OP components that
are compatible or incompatible with each other.  For a
tetragonal crystal of Hg1201 or NCCO, these OP components
act under the point group $C_{4v}$ according to Table I.
The OP components fall into sets, each of which has the
 same set of eigenvalues in the table.	Each set of OP
components with the same eigenvalues under the group operations forms a
 basis, from which an OP eigenfunction can be constructed.
We
have labeled these OP eigenfunctions as $|s\rangle$,
$|d_{x^2-y^2}\rangle$, $|d_{xy}\rangle$, $|g_{xy(x^2-y^2)}\rangle$, and
$|(p_x,p_y)\rangle$, respectively.
 We have also
listed the standard group theoretic (GT) notation for each OP basis
set. \cite{Tinkham}
The spin singlet OPs have one-dimensional representations, with eigenvalue
$+1$ under $E$.  Those labeled $A$ and $B$ are
even and
odd under $C_4$, respectively, and those with subscripts 1 and 2 are even and
odd under $\sigma_x$ and $\sigma_y$.
 Spin triplet OPs have two-dimensional
representations, with eigenvalue $+2$ under the identity operation $E$,
and  thus have GT
notation $E$. \cite{Tinkham,AGL,Machida}

But, how can we best represent the OP basis sets mathematically?
We shall employ two methods.
In the usual BCS theory, pairing occurs between otherwise free
quasiparticles	with opposite momenta $\pm {\bf k}_F$
on the Fermi surface.  For a free-particle
Fermi surface, the momenta are then described by the planar angle
$\phi_{\bf k}$, where ${\bf k}_F=k_F(\cos\phi_{\bf k}, \sin\phi_{\bf
k})$. The OP basis is then described
as in scattering theory, in which the scattered waves are
expanded in terms of angular momentum quantum numbers
$\ell$.     On the
other hand, for real-space pairing,  the pairing
takes place between quasiparticles on distinct lattice sites,
such as on
the same or nearest-neighbor sites. One then writes ${\bf
k}=(k_x,k_y)$, which components may satisfy a relation on the
real-space Fermi surface.  The first Brillouin zone (BZ) is given by
$|k_x|, |k_y|\le \pi/a$.
The real-space quantum numbers, $(n,m)$,  count the lattice
separations between the paired quasiparticles along the Cu-O bond
 directions.

In either scenario
there are an infinite number of possible OP basis elements,
since
$\ell$ is any integer satisfying $\ell\ge0$ and $n$, $m$ can be
 any integers.
However,  the crystal symmetry forces the
infinite number of OP basis elements to appear
together in a limited, finite number of OP
 basis {\it sets}, each set having an infinite number of elements.  We
thus describe each possible OP as having
 a set of  components, each component represented by a basis element.
Each  element in the OP basis set transforms identically under each of
the allowed crystal point group
operations.

 In the angular momentum $\ell$ representation, each OP
basis function (component) may be written as either $\cos(\ell\phi_{\bf k})$
 or $\sin(\ell\phi_{\bf k})$. In Fig. 2, we sketch
 the even $\ell$ OP basis functions  for
$\ell\le 4$
appropriate for singlet spin pairing.  Shaded and unshaded
 regions describe opposite signs of
the function.	 Shown are representations of the $\ell=0$
$s$-wave function, a constant, the two $\ell=2$ functions
 $d_{x^2-y^2}$ and $d_{xy}$, proportional to
 $\cos(2\phi_{\bf k})$ and $\sin(2\phi_{\bf k})$, respectively,
and the two $\ell=4$
functions
$g_{x^2y^2}$ and $g_{xy(x^2-y^2)}$, proportional to $\cos(4\phi_{\bf k})$ and
 $\sin(4\phi_{\bf k})$, respectively.

\begin{table}
\vbox{\tabskip=0pt
\def\tablerule{\noalign{\hrule}}
\halign to245pt{\strut#&#\tabskip=0.39em plus1em&
\hfil #& &\hfil#&#&\hfil#&#&\hfil#&#&
\hfil#&#&\hfil#&#&\hfil#&#\hfil\tabskip=0pt\cr
\noalign{\vskip5pt}\tablerule
\noalign{\vskip5pt} &&\omit\hidewidth GT \hidewidth&&
\omit\hidewidth OP\hidewidth&&
\omit\hidewidth $E$\hidewidth&&\omit\hidewidth $\sigma_{x}
$\hidewidth
&&\omit\hidewidth $\sigma_{d1}$\hidewidth&&\omit
\hidewidth $C_4$\hidewidth&&\omit\hidewidth $C_2$\hidewidth\cr
\noalign{\vskip5pt}
&&&&\hidewidth	Eigenfunction\hidewidth &&&&\omit
\hidewidth $\sigma_{y}$\hidewidth&&\omit\hidewidth $\sigma_{d2}$
\hidewidth&&\omit\hidewidth $C_4^{-1}$
\hidewidth
&&\cr
\noalign{\vskip5pt}
\tablerule \noalign{\vskip5pt}
 &&$A_1$&& $|s\rangle$&&+1&&+1&&+1&&
+1&&+1&\cr  &&$B_1$&&$|d_{x^2-y^2}\rangle$&&+1
&&+1&&
-1&&-1&&+1&\cr
&&$B_2$&&$|d_{xy}\rangle$&&+1&&-1&&
+1&&-1&&+1&\cr
&&$A_2$&&$|g_{xy(x^2-y^2)}\rangle$&&+1&&-1&&-1&&+1&&+1
&\cr
&&$E$&&$|(p_x, p_y)\rangle$&&+2&&0&&0&&0&
&-2&\cr
\noalign{\vskip5pt}\tablerule
\noalign{\vskip10pt}}}
\caption{Superconducting OP character table for a
tetragonal crystal such as Hg1201, with point group $C_{4v}$.	The symmetry
operations are the
identity ($E$), reflections
about the $x$ ($\sigma_x$), $y$ ($\sigma_y$), $d_1$
($\sigma_{d1}$), $d_2$ ($\sigma_{d2}$) axes, and rotations
about the
$c$-axis by $\pm 90^{\circ}$ ($C_4$, $C_4^{-1}$) and
180$^{\circ}$ ($C_2$).	The group theoretic
(GT) notations $A_1$, $A_2$, $B_1$, and $B_2$ are
one-dimensional representations, and $E$ is a
two-dimensional representation.}\label{t1}
\end{table}

The symmetry operations of group $C_{4v}$
(Table I) divide the
 even $\ell$ OPs into four basis sets, which are listed in Table II.
  These sets
consist of the	basis functions
$\cos(\ell\phi_{\bf k})$ and $\sin(\ell\phi_{\bf k})$, where
$\ell$ is even, and either is or is not divisible by 4, as indicated
in Table II.  Thus, the
$g_{x^2y^2}$ function, $\cos(4\phi_{\bf k})$ (equivalent to
$1-8\hat{k}_x^2\hat{k}_y^2$, which averages to zero), is an element of
 the same
OP basis set ($|s\rangle$) as the $\ell=0$, $s$-wave function, a
constant.
It is
possible to define  general $s$-wave eigenfunctions,
 $\Gamma_i(\phi_{\bf k})=\sum_{n=0}^{\infty}A_n^i
\cos(4n\phi_{\bf k})$, each of which transforms as the basis set denoted
$A_1$ in Table I.  We are then able to rewrite each of the OP
eigenfunctions in terms of the lowest $\ell$  basis functions
multiplying one of these generalized $s$-wave eigenfunctions,
as shown explicitly in Table II.

\begin{table}
\vbox{\tabskip=0pt
\def\tablerule{\noalign{\hrule}}
\halign to245pt{\strut#&#\tabskip=0.39em plus3em&
\hfil #& &\hfil#&#\hfil\tabskip=0pt\cr
\noalign{\vskip5pt}\tablerule
\noalign{\vskip5pt} &&\omit\hidewidth OP Eigenfunction\hidewidth&&
\omit\hidewidth    $\ell$ Representation
\hidewidth&\cr\noalign{\vskip5pt}
\tablerule
\noalign{\vskip5pt}
 &&\hbox{$|s\rangle$}&&\hbox{$\Gamma_1(\phi_{\bf k})$}&\cr\noalign{\vskip5pt}&&&&\hbox{$=a_0+
\sqrt{2}\sum_{n=1}^{\infty}a_n
\cos(4n\phi_{\bf k})$}&\cr\noalign{\vskip5pt}
&&\hbox{$|d_{x^2-y^2}\rangle$}&&\hbox{$\cos(2\phi_{\bf k})\Gamma_2(\phi_{\bf 
k})$}&\cr\noalign{\vskip5pt}&&&&
\hbox{$=\sqrt{2}\sum_{n=1}^{\infty}b_n\cos[(4n-2)
\phi_{\bf k}]$}&\cr\noalign{\vskip5pt}
&&\hbox{$|d_{xy}\rangle$}&&\hbox{$\sin(2\phi_{\bf k})
\Gamma_3(\phi_{\bf k})$}&\cr\noalign{\vskip5pt}&&&&\hbox{$=\sqrt{2}
\sum_{n=1}^{\infty}c_n
\sin[(4n-2)\phi_{\bf k}]$}&\cr \noalign{\vskip5pt} &&
\hbox{$|g_{xy(x^2-y^2)}\rangle$}&&\hbox{$\sin(4\phi_{\bf k})\Gamma_4(\phi_{\bf k})$}&\cr\noalign{\vskip5pt}&&&&\hbox{$=\sqrt{2}
\sum_{n=1}^{\infty}d_n
\sin(4n\phi_{\bf k})$}&\cr
\noalign{\vskip5pt}\tablerule
\noalign{\vskip10pt}}}
\caption{Angular momentum quantum number $\ell$ representations of the
singlet superconducting OP eigenfunctions for a tetragonal crystal.  The functions $\Gamma_i(\phi_{\bf 
k})=\sum_{n=0}^{\infty}A_n^i
\cos(4n\phi_{\bf k})$.}\label{t2}
\end{table}

In the lattice representation, the general tetragonal eigenfunctions are
 $\cos[a(nk_x+mk_y)]$ and $\sin[a(nk_x+mk_y)]$, where $n$
 and $m$ can be any integers.  One  classifies the
 OP  basis functions in terms of the spatial
separations of the paired quasiparticles.  In order of increasing
 separation, there is on-site (for a retarded BCS-type interaction),
near-neighbor, next-nearest-neighbor, etc. pairing.   Such expansions
 of the OP eigenfunctions are presented in Table III.
  In this representation, the most general OP
 eigenfunctions in the same basis set as the constant, $s$-wave OP
component can be written as $\tilde{\Gamma}_i({\bf k})=
\sum_{n,m=-\infty}^{\infty}\tilde{A}^i_{nm}\cos[a(nk_x+mk_y)]$,
 provided that we
choose the $\tilde{A}_{nm}^i$ to satisfy the $A_1$ ($s$-wave) symmetry,
 $\tilde{A}^i_{nm}=\tilde{A}^i_{-n,m}=\tilde{A}^i_{n,-m}=
\tilde{A}^i_{-n,-m}=\tilde{A}^{i}_{mn}$, of
 the $C_{4v}$ character table (Table I).  As for the
angular momentum representation, it is then possible to write each OP
 eigenfunction in terms of the nearest pairing separations
consistent with the representation group symmetry, multiplied by one of
the allowed generalized
$s$-wave  lattice eigenfunctions.

\begin{table}
\vbox{\tabskip=0pt
\def\tablerule{\noalign{\hrule}}
\halign to245pt{\strut#&#\tabskip=0.39em plus3em&
\hfil #& &\hfil#&#\hfil\tabskip=0pt\cr
\noalign{\vskip5pt}\tablerule
\noalign{\vskip5pt} &&\omit\hidewidth OP Eigenfunction\hidewidth&&
\omit\hidewidth    Crystal Representation
\hidewidth&\cr\noalign{\vskip5pt}
\tablerule
\noalign{\vskip5pt}
 &&\hbox{$|s\rangle$}&&\hbox{$\tilde{\Gamma}_1
({\bf k})$}&\cr\noalign{\vskip5pt}&&&&\hbox{$
=\sum_{n,m=0}^{\infty}a_{nm}[\cos(nk_xa)\cos(mk_ya)$}&\cr
\noalign{\vskip5pt}&&&&\hbox{$\>\>+\cos(mk_xa)
\cos(nk_ya)]$}&\cr
\noalign{\vskip5pt}
&&\hbox{$|d_{x^2-y^2}\rangle$}&&\hbox{$[\cos(k_{x}a)-
\cos(k_{y}a)]\tilde{\Gamma}_2({\bf k})$}\cr
\noalign{\vskip5pt}
&&&&\hbox{$ =\sum_{n,m=0}^{\infty}b_{nm}
[\cos(nk_xa)\cos(mk_ya)$}&\cr\noalign{\vskip5pt}&&&&
\hbox{$\>\>-\cos(mk_xa)\cos(nk_ya)]
$}&\cr
\noalign{\vskip5pt}
&&\hbox{$|d_{xy}\rangle$}&&\hbox{$\sin(k_{x}a)
\sin(k_{y}a)\tilde{\Gamma}_3({\bf k})$}&\cr
\noalign{\vskip5pt}
&&&&\hbox{$ =\sum_{n,m=1}^{\infty}c_{nm}[\sin(nk_xa)\sin(mk_ya)$}&\cr
\noalign{\vskip5pt}&&&&\hbox{$\>\>+
\sin(mk_xa)\sin(nk_ya)]$}&\cr\noalign{\vskip5pt}
 &&\hbox{$|g_{xy(x^2-y^2)}\rangle$}&&
\hbox{$\sin(k_{x}a)
\sin(k_{y}a)$}&\cr\noalign{\vskip5pt}&&&&
\hbox{$\times[\cos(k_{x}a)-\cos(k_{y}a)]
\tilde{\Gamma}_4({\bf k})$}&\cr\noalign{\vskip5pt}
&&&&\hbox{$ = \sum_{n,m=1}^{\infty}d_{nm}[\sin(nk_xa)\sin(mk_ya)
$}&\cr\noalign{\vskip5pt}&&&&
\hbox{$\>\>-\sin(mk_xa)\sin(nk_ya)]$}&\cr
\noalign{\vskip5pt}\tablerule
\noalign{\vskip10pt}}}
\caption{Lattice representations of the singlet
 superconducting OP
eigenfunctions
  for a
tetragonal crystal.
The functions $\tilde{\Gamma}_i({\bf k})=
\sum_{n,m=-\infty}^{\infty}\tilde{A}^i_{nm}\cos[a(nk_x+mk_y)]$, where
$\tilde{A}^i_{nm}=\tilde{A}^i_{-n,m}=\tilde{A}^i_{n,-m}=\tilde{A}^i_{-n,-m}
=\tilde{A}^{i}_{mn}$.}\label{t3}
\end{table}

Those OP
 components (or basis functions)
within the same basis set, which have the
same eigenvalues in the group character table, are
{\it compatible}, and can in principle mix freely,
the actual mixing amount depending upon the particular
 pairing interaction form.
OP components arising from different basis sets, with one or more
different eigenvalues in the group character table
are {\it incompatible}, and do not ordinarily mix.    Each OP
 eigenfunction is
 a linear combination of all of the compatible OP
basis functions  within the same basis set.  For tetragonal crystals,
 the
 OP eigenfunctions are listed in
Tables II and III.

In Table IV, we  list the spin-singlet
 character table and group theoretic notation
 for the  $C_{2v}$ group
operations for the orthorhombic crystal structure
appropriate for YBCO and Y124.	   \cite{Tinkham}
We also listed the angular momentum ($\ell$) and lattice ($n, m$)
 representations of the two OP eigenfunctions, $|s+d_{x^2-y^2}\rangle$
 and
$|d_{xy}+g_{xy(x^2-y^2)}\rangle$. Each of these OP eigenfunctions
  contains two
 basis sets  of the tetragonal OP eigenfunctions in Tables II or III.
Thus,
$|s+d_{x^2-y^2}\rangle$ contains an arbitrary mixing of the
 tetragonal $|s\rangle$ and $|d_{x^2-y^2}\rangle$ OP eigenfunctions,
 for example.

\begin{table}
\vbox{\tabskip=0pt
\def\tablerule{\noalign{\hrule}}
\halign to245pt{\strut#&#\tabskip=0.39em plus1em&
\hfil#& &\hfil#&#&\hfil#&#&\hfil#&#&\hfil#&#&
\hfil#&#
\hfil\tabskip=0pt\cr
\noalign{\vskip5pt}\tablerule
\noalign{\vskip5pt} &&\omit\hidewidth GT
\hidewidth&&
\omit\hidewidth OP \hidewidth&&\omit\hidewidth $E$
\hidewidth&&\omit\hidewidth $\sigma_x$
\hidewidth
&&\omit\hidewidth $\sigma_y$ \hidewidth&&
\omit\hidewidth $C_2$\hidewidth &\cr
\noalign{\vskip5pt}
&&&&\hidewidth Eigenfunction \hidewidth &&&
&\omit\hidewidth ($\sigma_a$)\hidewidth&&\omit
\hidewidth ($\sigma_b$)\hidewidth &&&\cr
\noalign{\vskip10pt}\tablerule \noalign{\vskip10pt}
 &&$A_1$&&\hbox{$|s+d_{x^2-y^2}\rangle$}&
&+1&&+1&&+1&&
+1&\cr\noalign{\vskip5pt}
&&&&\hbox{$ {\rightarrow\atop\ell}\>\> \tilde{a}_0+\sqrt{2}\sum_{n=1}^{\infty}[\tilde{a}_n\times$}
&&&&&&&&&\cr\noalign{\vskip5pt}
&&&&\hbox{$\qquad \times
\cos(2n\phi_{\bf k})]$}&&&&&&&&&\cr\noalign{\vskip5pt}
&&&&\hbox{$ {\rightarrow\atop{n,m}}\>\>
\sum_{n,m=0}^{\infty}[\tilde{a}_{nm}\times$}
&&&&&&&&&\cr\noalign{\vskip5pt}
&&&&\hbox{$\>\> \times\cos(nk_xa)\cos(mk_ya)]$}
&&&&&&&&&\cr\noalign{\vskip5pt}
&&$A_2$&&\hbox{$|d_{xy}+g_{xy(x^2-y^2)}\rangle$}
&&+1&&-1&&-1&&
+1&\cr\noalign{\vskip5pt}
&&&&\hbox{$ {\rightarrow\atop\ell}\>\> \sqrt{2}\sum_{n=1}^{\infty}[\tilde{b}_n\times$}
&&&&&&&&&\cr
\noalign{\vskip5pt}
&&&&\hbox{$ \qquad\times\sin(2n\phi_{\bf
k})]$}
&&&&&&&&&\cr
\noalign{\vskip5pt}
&&&&\hbox{$ {\rightarrow\atop{n,m}}\>\>
\sum_{n,m=1}^{\infty}[\tilde{b}_{nm}\times$}
&&&&&&&&&\cr
\noalign{\vskip5pt}
&&&&\hbox{$\>\>\times \sin(nk_xa)\sin(mk_ya)$]}
&&&&&&&&&\cr
 \noalign{\vskip5pt}\tablerule
\noalign{\vskip10pt}}}
\caption{Singlet superconducting OP eigenfunctions in the angular
momentum ($\ell$) and lattice ($n, m$) representations, their group
theoretic notations, and character table for the
orthorhombic point group $C_{2v}$ in the form
appropriate for YBCO. }\label{t4}
\end{table}

We remark that even with  only a slight (about 2\%) crystallographic
 difference between the $a$ and $b$ axes in YBCO, the
mixing of OP components that would have been
in different representations (or basis sets) of the tetragonal
group $C_{4v}$, can be very substantial.
For example, there could be a very large mixing of $s$-wave
and $d_{x^2-y^2}$-wave OP components in YBCO.
Especially as the penetration depth within the $ab$-planes
measured in YBCO is  anisotropic by about a factor of 2,
\cite{Bonn} and the $c$-axis Josephson tunneling between
untwinned YBCO and Pb  reproducibly gave values of
1.5 mV, \cite{Dynes} about 25\% of the expected result of
Ambegaokar-Baratoff, \cite{AB}, one expects the isotropic $\ell=0$
OP component  to be at least 25\% of the maximum OP amplitude
 at low $T$.  Of course, the penetration depth anisotropy is easiest
 to understand by accounting for the quasi-one-dimensional Fermi
 surface of the CuO chains.

\begin{table}
\vbox{\tabskip=0pt
\def\tablerule{\noalign{\hrule}}
\halign to245pt{\strut#&#\tabskip=0.39em plus1em&
\hfil#& &\hfil#&#&\hfil#&#&\hfil#&#&\hfil#&#&
\hfil#&#&\hfil#&#\hfil
\tabskip=0pt\cr
\noalign{\vskip5pt}\tablerule
\noalign{\vskip5pt}
 &&
\omit\hidewidth GT \hidewidth &&
\omit\hidewidth OP\hidewidth &&\omit\hidewidth $E$
 \hidewidth && \omit\hidewidth$\sigma_{d1}$\hidewidth
&&
\omit\hidewidth $\sigma_{d2}$\hidewidth
&&\omit\hidewidth $C_2$\hidewidth\cr
\noalign{\vskip5pt}
&&&&\hidewidth Eigenfunction \hidewidth&&&&
\omit\hidewidth ($\sigma_a$)\hidewidth &&\omit
\hidewidth ($\sigma_b$)\hidewidth &&
\cr
\noalign{\vskip5pt}\tablerule \noalign{\vskip5pt}
 &&$A_1$&&\hbox{$|s+d_{xy}\rangle$}
&&+1&&+1&&+1&&
+1&\cr\noalign{\vskip5pt}
&&&&\hbox{$ {\rightarrow\atop\ell}\>\>\tilde{a}_0+\sqrt{2}
\sum_{n=1}^{\infty}\{\tilde{a}_n\times$}&&&&&&&&&\cr
\noalign{\vskip5pt}
&&&&\hbox{$
\qquad\times\cos[2n(\phi_{\bf k}-\pi/4)]\}$}&&&&&&&&&\cr\noalign{\vskip5pt}
&&&&\hbox{$
{\rightarrow\atop{n,m}}\>\>\sum_{n,m=0}^{\infty}
\bigl\{[\tilde{a}_{nm}
+\tilde{a}_{mn}]\times$}&&&&&&&&&\cr\noalign{\vskip5pt}
&&&&\hbox{$\qquad\times\cos(nk_xa)\cos(mk_ya)$}
&&&&&&&&&\cr
\noalign{\vskip5pt}
&&&&\hbox{$
\qquad +[\tilde{c}_{nm}
+\tilde{c}_{mn}]\times$}&&&&&&&&&\cr\noalign{\vskip5pt}
&&&&\hbox{$\qquad\times\sin(nk_xa)\sin(mk_ya)\bigr\}$}
&&&&&&&&&\cr\noalign{\vskip5pt}
&&$A_2$&&\hbox{$|d_{x^2-y^2}+g_{xy(x^2-y^2)}
\rangle$}&&+1&&-1&&
-1&&+1&\cr\noalign{\vskip5pt}
&&&&\hbox{$ {\rightarrow\atop\ell}\>\> \sqrt{2}\sum_{n=1}^{\infty}\{\tilde{b}_n\times$}
&&&&&&&&&\cr
\noalign{\vskip5pt}
&&&&\hbox{$\qquad\times\sin[2n(\phi_{\bf
k}-\pi/4)]\}$}&&&&&&&&&\cr\noalign{\vskip5pt}
&&&&\hbox{$
{\rightarrow\atop{n,m}}\>\>\sum_{n,m=0}^{\infty}
\bigl\{[\tilde{a}_{nm}-\tilde{a}_{mn}]\times$}&&&&&&&&&\cr\noalign{\vskip5pt}
&&&&\hbox{$\qquad\times\cos(nk_xa)\cos(mk_ya)$}
&&&&&&&&&\cr
\noalign{\vskip5pt}
&&&&\hbox{$
\qquad +[\tilde{c}_{nm}-\tilde{c}_{mn}]\times$}&&&&&&&&&\cr\noalign{\vskip5pt}
&&&&\hbox{$\qquad\times\sin(nk_xa)\sin(mk_ya)\bigr\}$}
&&&&&&&&&\cr
 \noalign{\vskip5pt}\tablerule
\noalign{\vskip10pt}}}
\caption{Singlet superconducting OP eigenfunctions in the angular
momentum ($\ell$) and lattice ($n, m$) representations, their
 group
theoretic notations, and character table
for the orthorhombic point group $C_{2v}$ in the form
appropriate for
 BSCCO.   Although the
$\sigma_{d1}$
mirror plane symmetry is only approximate due to the
$c$-axis component of ${\bf Q}$, the $\sigma_{d2}$ mirror
plane
symmetry is robust in  most current samples.}\label{t5}
\end{table}

In Table V, we list the spin-singlet character table and group theoretic
notation for the
$C_{2v}$ orthorhombic group operations appropriate for BSCCO and LSCO.
We also listed the angular momentum ($\ell$) and lattice ($n, m$)
 representations of the two OP eigenfunctions, $|s+d_{xy}\rangle$
 and
$|d_{x^2-y^2}+g_{xy(x^2-y^2)}\rangle$. Each of these OP eigenfunctions
  contains two
 basis sets  of the tetragonal OP eigenfunctions in Tables II or III.
Thus,
$|s+d_{xy}\rangle$ contains an arbitrary mixing of the tetragonal  $|s\rangle$ and $|d_{xy}\rangle$ OP 
eigenfunctions, but
not of
the tetragonal $|d_{x^2-y^2}\rangle$ OP eigenfunction.
  We note that although the $ac$-plane
is not a strict BSCCO crystallographic mirror plane,	the $bc$-plane
is a strict crystallographic  mirror plane ($\sigma_{d2}$).  This fact is
sufficient to	separate
the OP sets as if the crystal were fully orthorhombic.

We remind the reader of the crucial difference between the OPs with the same
group theoretic notations for the orthorhombic symmetries
appropriate for YBCO
 and BSCCO, respectively.  Although for orthorhombic YBCO, OP components
that have $s$-wave and $d_{x^2-y^2}$-wave symmetry in the
tetragonal crystal can mix freely, and similarly OP components
exhibiting $d_{xy}$-wave and $g_{xy(x^2-y^2)}$-wave
 symmetry in the tetragonal crystal can mix, in BSCCO this is not the case.
Instead, the OP components exhibiting
$d_{x^2-y^2}$-wave and $g_{xy(x^2-y^2)}$-wave symmetry
in the tetragonal crystal can mix, as can the OP  components having
 $s$-wave and $d_{xy}$-wave symmetry in the tetragonal crystal representation.

\section{Ginzburg-Landau free energy}
In this section, we examine the Ginzburg-Landau free energy
of a superconductor with  one or more OP
components.   We limit our discussion
to a single CuO$_2$ layer, assuming equivalent OPs on all CuO$_2$
layers.
   We shall
see that there is a distinct difference
between the cases of two {\it compatible} OP components and
two {\it incompatible} OP
components.  (AGL used the
 terminology ``mixing'' and ``non-mixing'').

\subsection{single complex OP component}
We first assume a single OP
component.  Let the pairing interaction $\lambda({\bf k},{\bf
k}')=\lambda_0\varphi({\bf k})\varphi({\bf k}')$, where the
  basis function $\varphi({\bf k})$ is normalized,
 {\it i. e.,}  $\langle\varphi({\bf k})|\varphi({\bf
k})\rangle=1$, which in angular momentum representation implies
$\int_0^{2\pi}{{d\phi_{\bf k}}\over{2\pi}}
\varphi^2(\phi_{\bf k})=1$, or in real-space representation implies
$(a/2\pi)^2\int_{-\pi/a}^{\pi/a}dk_x\int_{-\pi/a}^{\pi/a}dk_y
\varphi^2({\bf k})=1$,
  and the OP eigenfunction
$\Delta({\bf k},T)=\Delta(T)\varphi({\bf k})$, where
$\Delta(T)$ is the OP amplitude.
The  Ginzburg-Landau free energy of a spatially and temporally uniform superconductor
with
a single OP component  is 
\begin{equation}
F=\alpha(T)|\Delta(T)|^2+\beta|\Delta(T)|^4,
\end{equation}
where $\alpha(T)=\alpha_0(T-T_c)$, $\beta > 0$ is a constant, and $T_c$ is obtained
from  $\lambda_0$ in the usual BCS
approximation.	Minimizing  $F$, we find $\Delta=0$ and $F_N=0$ for $T\ge T_c$, and
 $\Delta(T)=
\left[\alpha_0(T_c-T)/(2\beta)\right]^{1/2}$ and
 $F_S=-\alpha_0^2(T_c-T)^2/(4\beta)$ for $T<T_c$.

\subsection{Two incompatible OP components}

We now consider the case of two incompatible OP components.
For simplicity, we
 consider the case in which the pairing interaction can
be written in terms of  products of only two of the
incompatible, orthonormal basis functions $\varphi_i({\bf k})$ appropriate for
the crystal, $\lambda({\bf k},{\bf k}')=\sum_{i=1}^2
\lambda_{i0}
\varphi_i({\bf k})\varphi_i({\bf k}')$.
As examples of incompatible basis functions, in Hg1201, YBCO, or
BSCCO, we could choose the simplest $d_{xy}$-wave and
$d_{x^2-y^2}$-wave basis functions, which are $\sqrt{2}\sin(2\phi_{\bf
k})$ and $\sqrt{2}\cos(2\phi_{\bf k})$ in the $\ell$-representation.
For either Hg1201 or YBCO, we could also choose the simplest $s$-wave
and $d_{xy}$-wave functions (1 and $\sqrt{2}\sin(2\phi_{\bf k})$ in the
$\ell$-representation), and for either Hg1201 or BSCCO, we could also
choose the simplest $s$-wave and $d_{x^2-y^2}$-wave basis functions.
Of course,  the
number of possible incompatible OP component pairs
  is limitless.  The OP eigenfunction consistent with the above pairing interaction has the form
$\Delta({\bf k},T)=\sum_{i=1}^2
\Delta_i(T)\varphi_i({\bf k})$, where the
$\Delta_i(T)$ are complex constants representing the two incompatible
OP component amplitudes.

  For different $\lambda_{i0}$, the bare transition
temperatures $T_{ci}$ (obtained in the BCS model assuming
only one $\Delta_i\ne0$) of the two incompatible
components are different.  The
free energy can then be written as
\begin{eqnarray}
F&=&\sum_{i=1}^2\Bigl(\alpha_i(T)|\Delta_i|^2+
\beta_i|\Delta_i|^4\Bigr)\cr
& &\cr
& &
+\epsilon|\Delta_1|^2|\Delta_2|^2+
\delta[\Delta_1^2\Delta_2^{*2}
+\Delta_2^2\Delta_1^{*2}],\label{Fincompatible}
\end{eqnarray}
where the $\alpha_i(T)=\alpha_{i0}(T-T_{ci})$ and the
$\beta_i$, $\epsilon$ and $\delta$ are	constants.  In
weak
coupling (BCS) theory,  $\beta_i>0$, $\epsilon>0$, $\delta>0$, and
$\epsilon-2\delta>0$. Writing $\Delta_i=|\Delta_i|\exp(i\psi_i)$, we note that the
 only
term depending upon the phases $\psi_i$ is the one proportional
 to
$\delta$, which
 becomes $2\delta|\Delta_1|^2|\Delta_2|^2\cos[2(\psi_1-\psi_2)]$.
Assuming
 $T_{c1}>T_{c2}$, $F$ is minimized in the first superconducting (S$_1$) regime just
below $T_c$,  by
\begin{eqnarray}
|\Delta_1(T)|&=&[-\alpha_1(T)/(2\beta_1)]^{1/2},\cr
& &\cr
|\Delta_2|&=&0,\>\>\>
{\rm for}\> \>T_{c2}^{<}\le T< T_{c1}\>\> ({\rm S}_1).\label{S1}
\end{eqnarray}
In the low temperature (S$_2$) superconducting phase, both $|\Delta_1|\ne0$ and $|\Delta_2|\ne0$.   $\partial F/\partial\psi_i=0$ is satisfied when
 $\sin[2(\psi_1-\psi_2)]=0$, which minimizes $F$ when
$\psi_1-\psi_2=\pm\pi/2$.  The temperature $T_{c2}^{<}$ which separates the regions
S$_1$ from S$_2$ can then be obtained by inserting $|\Delta_1|$ from Eq. (\ref{S1}) into the linearized equation for $|\Delta_2|$, leading to
 \begin{eqnarray}
{{T_{c2}^{<}}\over{T_{c2}}}&=&{{1-\nu_1(T_{c1}/T_{c2})}\over{1-\nu_1}},\cr
& &\cr
\nu_1&=&{{\alpha_{10}(\epsilon-2\delta)}\over{2\alpha_{20}\beta_1}}.
\end{eqnarray}
Note that $T_{c2}^{<}<T_{c2}$ if $\epsilon-2\delta>0$, as in weak coupling (BCS) theory.  Then, solving a quadratic for $|\Delta_1|^2$ and $|\Delta_2|^2$
below $T_{c2}^{<}$, we obtain the solution in the low temperature
superconducting phase S$_2$,
\begin{eqnarray}
|\Delta_2|^2&=&{{\alpha_{20}(1-\nu_1)(T_{c2}^{<}-T)}
\over{2\beta_2(1-\nu_1\nu_2)}}\cr
& &\cr
|\Delta_1|^2&=&{{\alpha_{10}}\over{2\beta_1}}
\left[T_{c1}-T_{c2}^{<}-{{(1-\nu_2)(T_{c2}^{<}-T)}
\over{1-\nu_1\nu_2}}\right]\cr
& &\cr
\nu_2&=&{{\alpha_{20}(\epsilon-2\delta)}\over{2\alpha_{10}\beta_2}}\cr
& &\cr
\psi_2&=&\psi_1\pm\pi/2\>\>{\rm for}\>\> T<T_{c2}^{<} \>\>({\rm S}_2).
\end{eqnarray}
Because of the $\pi/2$ phase difference between the two OP components
in the S$_2$ state, this is commonly
known as the ``$\Delta_1+i\Delta_2$'' state.
As emphasized by AGL,  these two (or more) incompatible OP components
 can only be simultaneously non-vanishing below a second phase transition
 temperature, $T_{c2}^{<}$.  Since no
second phase transition appears to have been observed in
any of the cuprates (at least, not above 1 K as yet), it is
rather unlikely that $T_{c2}^{<}>1$ K.  In addition, the $\Delta_1+i\Delta_2$ states is nodeless below 
$T_{c2}^{<}$.

\subsection{Two compatible OP components}

Now we consider the more interesting case of two
compatible OP components, with  elements $\varphi_i({\bf k})$ for $i=1,
2$,  of the same basis set.
  In Hg1201, for example,  $|s\rangle$ contains the simplest $s$-wave
and $g_{x^2y^2}$-wave elements (written as $1$ and
$\sqrt{2}\cos(4\phi_{\bf k})$ in the $\ell$-representation), which are
compatible basis functions.  For YBCO, $\bigl|s+d_{x^2-y^2}\rangle$ contains
as compatible basis functions the simplest
$s$-wave and $d_{x^2-y^2}$-wave functions (1 and
$\sqrt{2}\cos(2\phi_{\bf k})$ in the $\ell$-representation), and for
BSCCO, the simplest $s$-wave and $d_{xy}$-wave functions are
compatible elements of  $\bigl|s+d_{xy}\rangle$.
 It is easy to generalize this procedure to include
all of the (infinite number of)
components in the relevant basis set.

 With two compatible OP components, the interaction is
generally
of the form $\lambda({\bf k},{\bf k}')=\sum_{i,j=1}^2
\varphi_i({\bf k})\lambda_{ij0}\varphi_j({\bf k}')$, where
$\tensor{\lambda_0}$ is a symmetric tensor of rank 2.  We
could use this form of $\lambda({\bf k},{\bf k}')$ to generate the
Ginzburg-Landau free energy,
letting
$\Delta({\bf k},T)=\sum_{i=1}^2
\Delta_i(T)\varphi_i({\bf k})$, where the
$\Delta_i$ are complex constants representing the OP
 component amplitudes.	We would then find that the free energy would
be given by Eq. (\ref{Fincompatible}), plus the terms
$\gamma
(\Delta_1^{*}\Delta_2+\Delta_1\Delta_2^{*})$ and
$(\Delta_1\Delta_2^{*}+\Delta_1^{*}\Delta_2)
(\mu_1|\Delta_1|^2
+\mu_2|\Delta_2|^2)$.\cite{AGL}  In the BCS approximation, the
$T_{ci}$ 
arise from the diagonal
elements $\lambda_{ii0}$, and $\gamma$ and the $\mu_i$ arise
from the  off-diagonal $\lambda_{120}=\lambda_{210}$ 
elements of $\tensor{\lambda}_0$, respectively.
One could then diagonalize the part of this free energy quadratic in
the $\Delta_i$ by a unitary transformation, a two-dimensional
rotation, as noted by AGL. \cite{AGL}  However, AGL did not emphasize
the important point that this rotation {\it mixes} the OP components
for all $T\le T_c$.  In order to clarify this point, we
choose instead to first diagonalize the pairing interaction. To do so,
we let $\tensor{\tilde\lambda}_0 =\tensor{R}\tensor{\lambda}_0\tensor{R}^{-1}$,
where $\tensor{R}$ is a standard two-dimensional matrix for rotation by
an angle $\theta$ about the $\hat{3}$ axis.  By choosing
$\tan(2\theta)=2\lambda_{120}/(\lambda_{110}-\lambda_{220})$, we 
obtain the diagonalized $\tensor{\tilde\lambda}_0$, and the
pairing interaction can now be written as $\lambda({\bf k},{\bf
k}')=\sum_{S=\pm}\varphi_S({\bf k})
\tilde{\lambda}_{S0}\varphi_S({\bf k}')$, where
\begin{eqnarray}
\varphi_{+}({\bf k})&=&\varphi_1({\bf k})\cos\theta+\varphi_2({\bf
k})\sin\theta\cr
& &\cr
\varphi_{-}({\bf k})&=&-\varphi_1({\bf k})\sin\theta+\varphi_2({\bf
k})\cos\theta,
\end{eqnarray}
and $\Delta({\bf k},T)=\sum_{S=\pm}\Delta_S(T)\varphi_S({\bf k})$.
We then expand the free energy in powers of $\Delta_{\pm}$, and obtain

\begin{eqnarray}
F& = &\sum_{S=\pm}\Bigl(\alpha_{S}(T)|\Delta_{S}|^2+\beta_{S}|
\Delta_{S}|^4\Bigr)\cr
& & +\epsilon|\Delta_{+}|^2|\Delta_{-}|^2 
+\delta(\Delta_{+}^2\Delta_{-}^{*2}+\Delta_{+}^{*2}
\Delta_{-}^2)\cr
& &\cr
& &+(\Delta_{+}\Delta_{-}^{*}+\Delta_{-}\Delta_{+}^{*})(\mu_{+}|\Delta_{+}|^2
+\mu_{-}|\Delta_{-}|^2),\label{Ftransformed}
\end{eqnarray}
where
$\alpha_{\pm}(T)= \alpha_{\pm0}(T-T_{c \pm})$, $T_{c\pm}$ is
obtained from
$\tilde{\lambda}_{\pm0}={1\over2}\{\lambda_{110}+\lambda_{220}
 \pm [(\lambda_{110}-\lambda_{220})^2+4\lambda_{210}^2]^{1/2}\}$
 in the BCS approximation.

Now, it is evident that the two original OP
components $\Delta_1$ and $\Delta_2$ mix in
 {\it two} ways.  First, and most important, the corresponding basis
functions $\varphi_i({\bf k})$
mix via the linear transformation employed to
diagonalize  the quadratic part of the free energy. Since the transformed OP
eigenfunctions are
$\Delta_{\pm}(T)\varphi_{\pm}({\bf k})$, 
{\it both} transformed OP eigenfunctions
contain contributions from {\it both} compatible basis functions.   Thus,
the dominant OP $\Delta_{+}$ (as well as the subdominant
$\Delta_{-}$) could be written as a ``$\Delta_1+\Delta_2$'' state.
 Although AGL did not emphasize it, this mixing
does {\it not} require a second phase transition.  

Second, the subdominant OP (with the lower
transition temperature, $T_{c-}$) is coupled to the dominant
OP just below $T_{c+}$ via the term $\mu_{+}|
\Delta_{+}|^2(\Delta_{+}\Delta_{-}^{*}+\Delta_{-}\Delta_{+}^{*})$.
Depending upon the sign of $\mu_{+}$, $\Delta_{-}$ will either
add or detract from $\Delta_{+}$, but in either case, it will be in
phase with it.	The most
important part of the resulting free energy just below $T_{c+}$
 is then
\begin{eqnarray}
F &\approx & -\alpha_{+}^2(T)/(4\beta_{+}) +
\alpha_{-}(T)|\Delta_{-}|^2\cr
& &\cr
& &
-2|\mu_{+}||\Delta_{-}|
\Bigl[-\alpha_{+}(T)/(2\beta_{+})\Bigr]^{3/2},
\label{Deltaminus}
\end{eqnarray}
which implies
\begin{equation}
|\Delta_{-}(T)| \approx {{2|\mu_{+}|}\over{\alpha_{-}(T_{c+})}}\Bigl({{-
\alpha_{+}(T)}\over{2\beta_{+}}}\Bigr)^{3/2},
\end{equation}
which behaves as $(T_{c+}-T)^{3/2}$.  Thus, the subdominant compatible
OP can pick up a strong temperature dependence just
below $T_{c+}$, modifying the relative temperature
dependences of $\Delta_1$ and $\Delta_2$ in the  ``$\Delta_1+\Delta_2$'' state.
 In particular, it is
possible
for one compatible component, say $\Delta_2$, to be
substantially smaller than the other just below $T_c$, but comparable
to  the other at low $T$.	 Again, this can
occur {\it without} a second phase transition, a point that was
incorrectly stated in the conclusion of AGL.

More generally,
when the (symmetric) interaction matrix $\tensor{\lambda}_0$ represents pairing involving
 a large (or infinite)
 number of compatible 
basis set elements, one diagonalizes it, and rank orders the diagonalized
interaction strengths $\tilde{\lambda}_{ii0}$.  Each eigenvalue
$\tilde{\lambda}_{ii0}$ corresponds to a bare $T_{ci}$ value, with
$T_c=T_{c1}$.  Each $\tilde{\lambda}_{ii0}$
 also corresponds to an OP eigenfunction which is 
a large (or infinite) sum of
the  compatible basis set elements.  The resulting free energy will
then contain quartic terms that mix two or more OPs.  These terms will
further mix the OPs, changing their effective $T$ dependences and
relative weights below $T_c$.  Thus,
the most general OP eigenfunction will have the general form given in
Tables II-V, with coefficients  (such as the $a_n$ for $|s\rangle$ in
Table II) having somewhat different $T$
dependences.  This implies that the ${\bf k}$-dependence of the OP eigenfunction changes smoothly with temperature. \cite{DESR}  {\it None} of this compatible 
mixing involves a second phase transition.

\section{$c$-axis twist Josephson junctions}
We now consider briefly the case of a HTSC Josephson junction formed
by twisting bicrystal halves an angle $\phi_0$ about the $c$-axis, as
pictured in Fig. 3. Let us
 suppose that there is only one phase transition observed in a
given material.
Certainly in the vicinity of the bulk transition $T_c$, it is hard
to imagine that a transition to a second state with a combined
OP that mixes incompatible components would
be present, except in the case of the triplet OPs.
However, since $|(p_x,p_y)\rangle$ is a two-dimensional representation
of the tetragonal group, it gives rise to a ``$p_x+ip_y$'' (or
equivalently, a ``$p_{d1}+ip_{d2}$'') state, as
in the equatorial plane of
 the $A$ phase of He$^3$.  In an orthorhombic crystal, the
transition temperatures $T_{c1}, T_{c2}$ for the two incompatible components are split, and a second
phase transition appears at $T{c2}^{<}$, below which the second component becomes
non-vanishing.  In any event, the combined OP eigenfunction will 
be
nodeless and rather isotropic at low $T$.
If this were true, one could not
presume that the ARPES, penetration depth, Raman, and
tunneling experiments were in any way probing the
superconducting OP, so that one would need to
explain those and related experiments from a model that did
{\it not} contain any nodes of the OP.

Thus, we assume that we are sufficiently close to
$T_c$ so that only one OP eigenfunction with compatible basis elements need
 be considered.    Let there be two crystal half-spaces, which
are rotated an angle $\phi_0$ with respect to each other.  On
the upper half-space, we let the OP be $\Delta_U(\phi_{\bf k}
-\phi_0/2)$, and on the lower
half-space, it is $\Delta_L(\phi_{{\bf k}'}+\phi_0/2)$.   We
assume that there is sufficient Josephson coupling across the
junction for a critical current to traverse it.

As we discussed in detail elsewhere, there are two  mechanisms
by which quasiparticles can tunnel across the twist
 boundary. \cite{KRS}  These are {\it coherent} and
{\it incoherent} tunneling, respectively.  In the coherent
tunneling process, the momenta parallel to the twist junction
are conserved during the tunneling, whereas for incoherent
 tunneling, they are ordinarily taken to be completely random.
We assume that the spatial average of the second order
quasiparticle tunneling processes across the twist boundary can
be written as
\begin{eqnarray}
\langle t({\bf k}-{\bf k}')t^{*}({\bf k}'-{\bf k}'')\rangle&=&
\delta({\bf k}-{\bf k}'')[|J|^2\delta({\bf k}-{\bf k}')\cr
& &\cr
& &\qquad
+f_{\rm inc}(\phi_{\bf k}-\phi_{{\bf k}'})],
\end{eqnarray}
where the overall two-dimensional $\delta({\bf k}-{\bf k}'')$
function insures translational invariance after averaging.  The
terms proportional to $|J|^2$  and $f_{\rm inc}(\phi_{\bf k}-\phi_{{\bf k}'})$ are the coherent and 
incoherent part of the
tunneling.  Ordinarily, one expects $f_{\rm inc}(\phi_{\bf k}-\phi_{{\bf k}'})$ to
be a constant, as for the standard $s$-wave scattering
calculation of Ambegaokar-Baratoff. \cite{AB}
  However, to allow for a small amount of forward scattering, or
incoherent tunneling of non-$s$-wave character, we let
\begin{equation}
f_{\rm inc}(\phi_{\bf k}-\phi_{{\bf k}'})=
\sum_{\ell=0}^{\infty}{{\cos[\ell(\phi_{\bf k}-\phi_{{\bf k}'})]}\over{2\pi\tau_{\perp\ell}N_{2D}(0)}},
\end{equation}
where $N_{2D}(0)=m/(2\pi)$ is the two-dimensional single-spin
quasiparticle density of states. \cite{KRS}

We remark that the question of coherent versus incoherent
tunneling is an interesting one, especially when one considers tunneling
 between two layers that are twisted about the
$c$-axis with respect to each other.   The important point is that a
quasiparticle tunnels from the Fermi surface on one layer to the Fermi
 surface on the neighboring layer.  This is true  in both the normal and
superconducting states.  For the simple case of a
quasi-two-dimensional free-particle band, the
cross-section of the Fermi surface in the $k_x/k_y$ plane is a
circle centered
 about the high-symmetry $\Gamma$ point at the center of
the first BZ.	 Even for
the case of layers twisted  about the $c$-axis, the circular Fermi
surface cross-sections are
identical in both first BZs twisted an angle $\phi_0$ with respect to
each other. In the coherent single particle
tunneling process, one can in principle tunnel from one layer
to the next from any position within the first BZ, and for such a
free particle Fermi surface, coherent normal state quasiparticle tunneling
could occur with equal probability over this circular cross-section,
even for arbitrary $\phi_0$, as long as the applied voltage $V$ was
effectively zero.

In the cuprates, however, the CuO$_2$ bands  generally have the
tight-binding, rather than the free-particle form.  For BSCCO, the
actual Fermi surface is rather complicated, as
pictured in Fig. 4a.  First of all, the
primary
Fermi surface is
very similar to the tight-binding Fermi surface indicated by
the thick curves in Fig. 4a.  These curves were calculated
using the two-dimensional tight-binding dispersion
\begin{eqnarray}
\xi({\bf k})&=&t[\cos(k_xa)+\cos(k_ya)]\cr
& &\cr
& &
-t'\cos(k_xa)\cos(k_ya)-\mu,
\end{eqnarray}
where the primary Fermi surface is obtain by setting $\xi({\bf k}_F)=0$.
 In Fig. 4a, we used the values $t=2t'=0.8\mu$,
which
gives a primary Fermi surface similar to that calculated for
 BSCCO and observed in ARPES experiments.
\cite{Ding,Freeman,Olson}
  In addition, there are the secondary Fermi surfaces, which
arise from the periodic lattice distortion ${\bf Q}=(0,0.212,1)$
in terms of the reciprocal lattice vectors.
\cite{Moss1}  These are given by $\xi({\bf k}_F\pm{\bf Q})=0$, which
are indicated by the thin solid and dotted curves in Fig. 4a.
Note that
for this two-dimensional dispersion, the $c$-component of
${\bf Q}$ is irrelevant to the Fermi surface structure.  For
$c$-axis tunneling between adjacent layers that lie directly
on top of one another, one could in principle still have coherent tunneling
 at any position on the multiple Fermi surfaces within the
first BZ.  When adjacent layers are twisted an angle $\phi_0$
about the $c$-axis with respect to each other, as pictured for
$\phi_0=\pi/4$ in Fig. 4b, however, the Fermi surfaces on
adjacent layers intersect only at a finite set of points.  This
 intersection has
measure zero relative to the entire length of the
two-dimensional Fermi surface. We thus expect the amount of
coherent tunneling between layers twisted by a sizable
 angle $\phi_0$ to be vanishingly small.  Otherwise,
preserving
 the wavevector during the coherent tunneling process
necessarily requires inelastic processes.  Thus, except for
the cases of $\phi_0\approx0,\pi/2$, we expect the
dominant
interlayer tunneling processes to be {\it incoherent}.

 Previously, \cite{KRS} we  treated the theoretically more 
interesting (and
more complicated) case in which the composition of the OP components can
depend upon the layer index, as the twist junction specifically
removes translational invariance.  For the case in which the
dominant OP component is presumed to be  $d_{x^2-y^2}$-wave 
and the
secondary, incompatible OP component is  $d_{xy}$-wave,  the dominant
OP component
is suppressed by the proximity to the
twist junction, locally reducing 
the suppression of the sub-dominant OP component 
in the bulk of the sample 
 by the	dominant OP component.
   In addition, for all $T<T_c$,  this sub-dominant OP
component  would be locally enhanced, although the enhancement would
 become quite weak close to $T_c$.  This would
allow the local OP to twist, compensating for the
physical
twist in the junction, and
allowing a finite amount of Josephson critical current at all
twist angles.

We found, however, that sufficiently close to the bulk $T_c$,
only one OP component need be considered, except
for the special (and only infinitesimally probable) case of an
 {\it accidental degeneracy}, whereby two  incompatible OPs
accidentally happen to have (almost) exactly the same bare
 transition temperature, and combine to give a nodeless
$\Delta_1+i\Delta_2$ state. \cite{KRS}  Otherwise, assuming the 
bare
$T_{cB}$ of the secondary $d_{xy}$ OP component is sufficiently small
that $T_{cB}^{<}$, the suppressed secondary transition in the bulk,  
is unobservable experimentally, the Josephson coupling strength
determines the $T$ at which the OP twisting could occur.  The weaker
the coupling, the lower the $T$ at which the OP can compensate for the
junction twist by OP twisting, and the experiment becomes 
definitive
in the vicinity of $T_c$.  Such effects were shown in Fig. 4b of our
earlier paper. \cite{KRS}  In that case, we showed explicitly that
for  strong ($\eta_d=\eta_d^{'}=1$)
 $d$-wave Josephson coupling across the twist junction 
($\eta_d^{'}$) and
non-junction ($\eta_d$) layers, respectively, a small but finite
$I_c(\phi_0=45^{\circ})$ was obtained at $t=T/T_{cA}=0.7$.  However,
reducing $\eta^{'}_d$ to 0.01 dramatically reduced
$I_c(\phi_0=45^{\circ})$ at $t=0.7$.  Moreover, as the experiments 
of
Li {\it et al.} now indicate that $\eta=\eta'<<1$ in BSCCO, we have
recalculated $I_c(\phi_0)$ for the cases $\eta_d=\eta_d^{'}=0.1, 0.001$, and
plotted the results for $t=0.5, 0.9, 0.99$ in Fig. 5.  The
remaining parameters are the same as before, and are listed in 
the
caption to Fig. 5.  It is seen that for $T_{cB}/T_{cA}=0.2$ (about 
as
high as one could possibly hope to have, in order for quasi-nodes 
to
still be present at 1 K), both sets of curves show the suppression 
of
the dominant OP component near to $T_c$, with 
$I_c(\phi_0)/I_c(0)$
lying below $|\cos(2\phi_0)|$ except for 
$\phi_0\approx45^{\circ}$.
They also show that 
 $I_c(\phi_0=45^{\circ})$
decreases with increasing $T/T_{cA}$ and decreasing $\eta_d=\eta_d^{'}$.
Clearly, twisting of the purported $d$-wave OP  {\it cannot} explain the
experiment of Li {\it et al.}, in which $I_c(\phi_0)$ was found to be
independent of $\phi_0$.

We therefore now consider the case in which there is only one OP, but
that it may contain a complex mixture of compatible basis set functions.
In the Ginzburg-Landau regime near to $T_c$, we found that the Josephson
critical current across the twist junction could be written in
terms of coherent and incoherent parts,  $I_c^{\rm coh}(\phi_0,T)
+I_c^{\rm inc}(\phi_0,T)$, \cite{KRS} where
\begin{eqnarray}
I^{\rm coh}_c(\phi_0,T)&\rightarrow&2eN_{2D}(0)b_0(T)|J|^2
{\rm Im}\!\!\int_0^{2\pi}\! \! d\phi_{\bf k}\cr
& &\cr
& &\>\>
\times\Delta_U(\phi_{\bf k}-\phi_0/2)\Delta_L^{*}(\phi_{\bf k}
+\phi_0/2)\cr
& &\cr
\noalign{\hbox{\rm and}}
& &\cr
I_c^{\rm inc}(\phi_0,T)&\rightarrow&eN^2_{2D}(0)a_0(T)
{\rm Im}\!\!\int_0^{2\pi}\!\! d\phi_{\bf k}\! \!\int_0^{2\pi}\!\!
d\phi_{{\bf k}'}\cr
& &\cr
& &\>\>\times\Delta_U(\phi_{\bf k}-\phi_0/2)f_{\rm inc}
(\phi_{\bf k}-\phi_{{\bf k}'})\cr
& &\cr
& &\qquad\times\Delta_L^{*}(\phi_{{\bf k}'}+
\phi_0/2),
\end{eqnarray} where $a_0(T)=\pi/(4T)$ and $b_0(T)=7\zeta(3)/(8\pi^2T^2)$.
Near $T_c$, 
 $I_c(\phi_0,T)$ is  proportional to $(T_c-T)$.

We then have to evaluate $I_c(\phi_0)$ for the various
OP possibilities for Hg1201, YBCO, and BSCCO.  To
do so, it is convenient to define
\begin{eqnarray}
\eta_0^{\rm inc}&=&e\pi N_{2D}(0)a_0(T)/\tau_{\perp0}\cr
& &\cr
\tilde{\eta}_{\ell}&=&{{\tau_{\perp 0}}\over{\tau_{\perp \ell}}}+
{{2b_0(T)|J|^2\tau_{\perp 0}}\over{a_0(T)}}.
\end{eqnarray}
Clearly, $\eta_0^{\rm inc}$ is	one-half the amplitude of the
 incoherent scattering contribution to the critical current for the
$s$-wave OP component, and $\tilde{\eta}_{\ell}$ is a measure
of the coherent and incoherent contributions to $I_c$ from the
$\ell$-wave OP components, relative to that of 
$\eta_0^{\rm inc}$.  Note that $\tilde{\eta}_2\propto\eta_d^{'}$ 
in
 Fig. 5.

\begin{table}
\vbox{\tabskip=0pt
\def\tablerule{\noalign{\hrule}}
\halign to245pt{\strut#&#\tabskip=0.39em plus3em&
\hfil#& &\hfil#&#&\hfil#&#\hfil\tabskip=0pt\cr
\noalign{\vskip5pt}\tablerule
\noalign{\vskip5pt} &&\omit\hidewidth GT \hidewidth&&
\omit\hidewidth OP \hidewidth &&
\omit\hidewidth $I_c(\phi_0)/\eta_0^{\rm inc}$\cr
\noalign{\vskip5pt}\tablerule \noalign{\vskip5pt}
 &&$A_1$&&$|s\rangle$&&\hbox{$\Bigl|a_0^2+
\sum_{n=0}^{\infty}
\tilde{\eta}_{4n}a_n^2\cos(4n\phi_0)\Bigr|$}&\cr
\noalign{\vskip5pt}
&&$B_1$&&$|d_{x^2-y^2}\rangle$&&\hbox{$\Bigl|\sum_{n=1}^{\infty}
\tilde{\eta}_{4n-2}b_n^2\cos[(4n-2)\phi_0]\Bigr|$}&\cr
\noalign{\vskip5pt}
&&$B_2$&&$|d_{xy}\rangle$&&\hbox{$\Bigl|
\sum_{n=1}^{\infty}
\tilde{\eta}_{4n-2}c_n^2\cos[(4n-2)\phi_0]\Bigr|$}&\cr
\noalign{\vskip5pt}&&$A_2$&&$|g_{xy(x^2-y^2)}\rangle$
&&\hbox{$\Bigl|
\sum_{n=1}^{\infty}\tilde{\eta}_{4n}d_n^2\cos(4n\phi_0)
\Bigr|$}&\cr
\noalign{\vskip5pt}\tablerule
\noalign{\vskip10pt}}}
\caption{Superconducting critical current across junctions twisted by
 $\phi_0$  about the $c$-axis of tetragonal crystals such as Hg1201 (from Table II).}\label{t6}
\end{table}

In Table
VI, we list the resulting $I_c(\phi_0)/\eta_0^{\rm inc}$ for the 
singlet OPs in
 tetragonal materials such as Hg1201.
 In
Table VII, $I_c(\phi_0)/\eta_0^{\rm inc}$
is presented
for the orthorhombic materials BSCCO and
YBCO.  We note, however, that in this case the actual forms of
the OP eigenfunctions involved in the respective states of
those two orthorhombic materials are different, as they are
given in Tables
IV and V.   From Table VI,  the  generalized
$d$-wave states (with
$B$ group theoretic symmetry) are 
 indistinguishable from each other. The leading $\ell=2$ components of
these generalized $d$-wave states give rise to  $I_c(\phi_0=45^{\circ})=0$.  The leading $\ell=4$ 
contribution to
the generalized $g_{xy(x^2-y^2)}$-wave
state likewise give rise to $I_c=0$ at $\phi_0=\pm 22.5^{\circ}$ and $\pm
67.5^{\circ}$.	For completeness, the leading
$\ell=1$ triplet OP components give rise to
$I_c(\phi_0=90^{\circ})=0$.  These simple leading $\ell$ cases give rise to
$I_c(\phi_0)\propto\bigl|\cos(\ell\phi_0)\bigr|$.
These simple cases are pictured in Fig. 6.

Including compatible components of higher $\ell$ values
will not change these qualitative results.
 In addition, if the tunneling across the
twist junction is entirely incoherent, with $s$-wave
 incoherent tunneling only ($\tau_{\perp\ell}
\rightarrow\infty$
for $\ell\ne0$), then these states will give a vanishing
critical current for all twist angles $\phi_0$ (even for
 the untwisted
case $\phi_0=0$), and the generalized $s$-wave state
would result in a constant $I_c(\phi_0)$, the constant
being a measure of the $\ell=0$, pure $s$-wave part of the
  OP eigenfunction.

On the other hand, if the relative contribution of the pure
$s$-wave form to the entire OP eigenfunction were very
small ({\it i. e.}, nearly vanishing), then a more complicated
scenario could develop.  For
instance, if either higher $\ell$-incoherent or
coherent intertwist tunneling were present and substantial,
and if the
contribution to the intertwist tunneling arising from the
$g_{x^2y^2}$ component
were larger than the contribution arising from
the $s$ component, and the $\ell\ge 8$-wave components
of the OP vanished, then there could be angles
$\phi_0^{*}$ satisfying $\phi_0^{*}={1\over4}\cos^{-1}
[-a_0^2(1+\tilde{\eta}_0)/\tilde{\eta}_4a_1^2]$ at
which $I_c(\phi_0^{*})=0$.  Including higher $\ell$
components of the generalized $s$-wave OP could
cause
the actual $\phi_0^{*}$ values to deviate from this simple
formula, however, as pictured in Figs. 7a, 7b, and 7c.	Such a
scenario is mainly expected in the
case of a nearly vanishing $\ell=0$ OP component, since the
 tunneling at a finite twist angle is most likely incoherent for
 tight-binding Fermi surfaces.

\begin{table}
\vbox{\tabskip=0pt
\def\tablerule{\noalign{\hrule}}
\halign to245pt{\strut#&#\tabskip=0.39em plus3em&
\hfil#& &\hfil#&#\hfil\tabskip=0pt\cr
\noalign{\vskip5pt}\tablerule
\noalign{\vskip5pt} &&\omit\hidewidth	GT \hidewidth&&
\omit\hidewidth $I_c(\phi_0)/\eta_0^{\rm inc}$\cr
\noalign{\vskip5pt}\tablerule \noalign{\vskip5pt}
&&$A_1$&&\hbox{$\Bigl|\tilde{a}_0^2(1+
\tilde{\eta}_0)+
\sum_{n=1}^{\infty}\tilde{\eta}_{2n}\tilde{a}_n^2
\cos(2n\phi_0)
\Bigr|$}&\cr\noalign{\vskip5pt}
&&$A_2$&&\hbox{$\Bigl|
\sum_{n=1}^{\infty}\tilde{\eta}_{2n}\tilde{b}_n^2\cos(2n\phi_0)
\Bigr|$}&\cr
 \noalign{\vskip5pt}\tablerule
\noalign{\vskip10pt}}}
\caption{Superconducting critical current across $c$-axis
junctions twisted by $\phi_0$, for orthorhombic YBCO or BSCCO, from
Tables IV and V.  See text.}\label{t7}
\end{table}

For an orthorhombic triplet superconductor, a twist angle of $\pm 90^{\circ}$
should
give rise to a vanishing critical current, as pictured in Fig. 6.
This is clearly
contradicted by the present experiments on BSCCO
of Li {\it et al.},
which show that a twist angle of 90$^{\circ}$ is
indistinguishable from an untwisted junction. \cite{Li2,Li3}
Thus, it is tempting to use this result to rule out the triplet OPs.
However, as noted above, the near degeneracy of the 
$p_{d1}$-wave and
$p_{d2}$-wave OPs in BSCCO could allow the dominant OP to 
twist,
giving rise to a very small, but nearly isotropic $I_c(\phi_0)$.
Thus, we cannot rule out the triplet states without some 
additional
 information.  As mentioned above, however, such triplet 
states are
expected to be nodeless for most of the regime $T<T_c$.

Now, the more interesting cases are the singlet spin states.
There are two possibilities, states with group theoretic
notation $A_1$ and $A_2$.  In Figs. 7a, 7b, and 7c, we have 
pictured
some examples of the various possibilities.  In the $A_2$
 case, there is no $s$-wave compatible
component,
so there are quite generally twist angles $\phi_0^{*}$ at
which the critical current vanishes.  For the pure 
$d_{xy}$-wave
OP for YBCO, or a pure  $d_{x^2-y^2}$-wave OP
for BSCCO,  $I_c(\phi_0^{*}=\pm45^{\circ})=0$, as illustrated 
in Fig. 6.
Similarly, for the
 OP  case with GT notation $A_2$, of a pure 
$g_{xy(x^2-y^2)}$-wave OP, 
$I_c(\phi_0^{*})=0$ at $\phi_0^{*}=\pm 22.5^{\circ},
\pm 67.5^{\circ}$, as shown in Fig. 6.
 More generally, we then expect that there will be twist
angles	between these two limiting cases at which the
$c$-axis critical current will vanish, as illustrated in Figs. 7a,
7b, and 7c.  For example, if  only $\tilde{b}_1$
and
$\tilde{b}_2$ are non-vanishing, then the critical current
 vanishes at
\begin{equation}
\phi_0^{*}={1\over2}\cos^{-1}\left({{-1+[1+8z^2]^{1/2}}\over{4z}}\right),
\end{equation}
where $z=(\tilde{\eta}_4/\tilde{\eta}_2)(\tilde{b}_2/\tilde{b}_1)^2$.
This expression would be modified
somewhat by including $\ell=6$ and higher order
components, but the results would still show a strong
anisotropy of the critical current,   with some
angles $\phi_0^{*}$ at which $I_c(\phi_0^{*})=0$.  Of course, if
the intertwist tunneling were only incoherent, but contained a small amount
of $d$-wave incoherent tunneling ($\tau_{\perp2}<\infty$), but no
 higher-$\ell$-wave incoherent tunneling, then	$\phi_0^{*}=\pm45^{\circ}$.

As indicated in Fig. 7a, the  $A_1$ state, which
contains the $s$-wave component,  can give the least
anisotropic (and non-vanishing) $I_c(\phi_0)$ behavior.  However, if
the $s$-wave
component is very small (Fig. 7b) or vanishes (Fig. 7c), then a highly
anisotropic $I_c(\phi_0)$ can occur.  In YBCO, most
workers apparently believe that the OP
is
$|s+d_{x^2-y^2}\rangle$, which contains  $d_{x^2-y^2}$-wave
and $s$-wave OP components.	Although the
experiment has not
yet been
performed in YBCO, this simple analysis could give a  measure
of the relative
mixing of $s$-wave and $d_{x^2-y^2}$-wave OP components in
YBCO, as a function of $T$.

In addition, this experiment would allow for a measurement of the
 relative weight of coherent to incoherent tunneling, both of which are
thought to be present in overdoped YBCO, at least.  Since
coherent tunneling is expected to occur only for $\phi_0=0,\pi/2$,
measuring $I_c(\phi_0)$ for $\phi_0$ near to these values could
 give valuable information  in this regard.  For $\phi_0$ values
away from $0,\pi/2$, one could measure the relative importance
of the $s$-wave and $d$-wave parts of the incoherent tunneling,
along with the relative weight of each OP component as a function
 of $T$.
In particular, this  would give a measure of the lower limit of the amplitude
 of the $s$-wave component of the OP.  To date, no such lower
limit has been placed in the vicinity of $T_c$ for YBCO.

In BSCCO, the situation
is fundamentally different, as the $d_{x^2-y^2}$-wave
OP is the leading component of the $A_2$ state
rather than the $A_1$ state.  Thus, if the present experiments
of Li {\it et al.} stand the test of time, the lack of any
$\phi_0$ dependence of $I_c$ across the BSCCO $c$-axis twist
junctions is {\it prima facie} evidence that the dominant OP
 is  $|s+d_{xy}\rangle$,
which
does {\it not}	contain any component of the purported
$d_{x^2-y^2}$-wave
 symmetry.  Furthermore, it would be strong evidence   that a
$d_{x^2-y^2}$-wave OP component (as a  part of the incompatible 
$|d_{x^2-y^2}+g_{xy(x^2-y^2)}\rangle$ OP) could
{\it only}
 appear below a second  phase transition.  Especially if one were to imagine
that the purported $d_{x^2-y^2}$-wave OP component were to
be larger than the $s$-wave component, then one would require
 a second specific heat peak {\it at least} as large as that
observed at $T_c$.
In addition, the incompatibility of the $s$-wave and
$d_{x^2-y^2}$-wave OP components would mean that the OP
containing a purported $d_{x^2-y^2}$-wave component
would be of the ``$\Delta_1+i\Delta_2$'' type, which would
{\it not} have any nodes below the second transition.

We remark that the apparent absence of any $\phi_0$
dependence of $I_c$ in the vicinity of $T_c$ does not
necessarily mean that the OP is pure $s$-wave.	It only
proves that the OP is 
 $|s+d_{xy}\rangle$ (Table V).  If the intertwist
tunneling were pure $s$-wave incoherent tunneling, one
could get a
$\phi_0$-independent $I_c$ for any $|s+d_{xy}\rangle$ basis set
combination, as long as the $s$-wave, $\ell=0$ OP component
 is finite.  That is,
the $d_{xy}$-wave or $g_{x^2y^2}$-wave OP component  could even
 be much larger than the $s$-wave OP component, giving rise to
nodes in the OP,
and one could still observe a $\phi_0$-independent $I_c$.
However, symmetry still requires that a $\phi_0$-independent
$I_c(\phi_0)$   {\it precludes} the simultaneous presence
 of any $d_{x^2-y^2}$-wave OP component.	 An
example of an OP that would be roughly
 consistent with the ${\bf k}$-dependence
 of the quasiparticle gap observed in ARPES
experiments (and also consistent with the $c$-axis twist
experiments of Li {\it et al.})
would be the so-called ``extended $s$-wave'' OP,
proportional to $|\cos(2\phi_{\bf k})|$, which is a specific
 example of $|s\rangle$ listed in Table II.  This OP has
nodes, but no $d_{x^2-y^2}$-wave component.

We
remark that  it is really possible to do these
experiments as a function of $T$, measuring
$I_c(T,\phi_0)$. \cite{Li3} Since the lack of $\phi_0$
dependence observed just below $T_c$ persists
to low $T$ values in all samples for which it can be measured both in
the single crystal and across the twist junction, then it most likely
 that the OP is {\it entirely}
$|s+d_{xy}\rangle$, without
any $d_{x^2-y^2}$-wave component at any measurable
temperature.  Such a scenario would be 
consistent with the  Pb/BSCCO $c$-axis
Josephson tunneling results. \cite{Kleiner}
 Although those authors observed
 very small $I_cR_n$ values, which could by themselves be interpreted
 as evidence for a very small $s$-wave
component to the OP, other possible, more mundane
explanations of such small values are easy to imagine, even
if the superconducting OP were essentially pure $s$-wave.
\cite{LK}  For example, the same low $I_cR_n$ values were 
observed in
NCCO, \cite{Woods} which suggests that problems of a materials nature,
such as an impedance mismatch in $c$-axis tunneling between cuprates
and Pb, might  be responsible for reducing  $I_cR_n$.

\section{Improved tricrystal experiment proposal}

We note that the tricrystal experiments of Kirtley, Tsuei,
and collaborators have not been confirmed in a second  
laboratory.
This could be due to the fact that it is very expensive
to prepare the SrTiO$_3$ tricrystal substrates upon which
the cuprate films were grown.  Such depositions
led to many complications at the grain boundaries, as have
 been evidenced by detailed transmission electron microscopy
(TEM) in the case of YBCO junctions. \cite{Miller} However,
in the cases of
other cuprates there have not been any TEM measurements
available to the general public.  Presumably, this is due to
the large substrate costs involved. 

We thus propose a new procedure, which does {\it not} employ
any expensive tricrystal substrate, but only requires an inexpensive
 insulator to support
the
tricrystal.
  The
improved geometry for the ``tricrystal'' experiment is actually
 constructed out of a tetracrystal ring, as pictured in Fig. 8.
In this setup, a single crystal of BSCCO is cleaved twice, giving
three pieces, which we illustrated as  dark,  intermediate,
 and light.  The intermediate  piece
is cut normal to the layers into two pieces of identical
thickness.  The dark crystal  is placed upon the support.
A section of support substrate of thickness equal to that of
the dark  crystal  is placed aside it
as a support for the rest of the ring.	Then, the two pieces of
the intermediate crystal  are placed across the dark crystal and the support,
forming angles $\phi_{12}$ and $\phi_{23}$, as pictured.  The
two pieces of the intermediate crystal	do not touch each other.  Instead, the
connection between them is made by placing the light crystal  on 
top
of  both intermediate crystals,
 forming a straight (0$^{\circ}$ or 180$^{\circ}$)
angle with one
of them.  Then, the entire triangular ring is fused together just
below the melting point, and mounted on the insulating support.   The experiment to test for the
flux trapped within the tricrystal can now be performed, whether 
by
measuring the flux within the tetracrystal with a SQUID
microscope, or by applying leads, and measuring the
tetracrystal SQUID characteristics.  In scanning with a SQUID microscope, one likely needs to fill the 
central region of the tetracrystal with an insulating material such as epoxy, and to make the triangular 
central region small enough to enhance 
the microscope sensitivity.

Let us label the crystals (1), (2), (3), and (4), clockwise
beginning with the intermediately-shaded crystal in the
 lower right-hand
part
of Fig. 8. Then, the angles of the junctions as one goes 
around the tetracrystal are $\phi_{12}$, $\phi_{23}$,
$\phi_{34}=0^{\circ}$
or $180^{\circ}$, and $\phi_{41}=\pm\phi_{31}$.  For
each of the four junctions in this tetracrystal, the critical
current can be calculated for the OP eigenfunction appropriate
for the
BSCCO crystal symmetry, taking into account the Josephson
 coupling
strengths.
If the
$s$-wave OP component dominates, as in the experiment of 
Li {\it et al.},
$I_c$ would
 be predicted to have the same sign across each of the
junctions, so
that the number of $\pi$-junctions would be 0.  However, if
the $d_{x^2-y^2}$ OP
component dominates, then the critical currents behave as
 $I_{c12}\cos(2\phi_{12})$, $I_{c23}\cos(2\phi_{23})$, $\pm I_{c34}$,
and $\pm I_{c41}\cos[2(\phi_{12}+\phi_{23})]$, where the
$\pm$
are consistent, and relate to whether it is a $0^{\circ}$ or
180$^{\circ}$ junction.  For choices of the angles such that
$\cos(2\phi_{12})$, $\cos(2\phi_{23})$ and
$\cos[2(\phi_{12}+
\phi_{23})]$ are all $\le0$, then it is easy to show for
the
16 possible configurations that one always has an odd
number
 of $\pi$-junctions.  Thus, if the
 $d_{x^2-y^2}$-wave OP component dominated the $c$-axis Josephson
tunneling, as required for a bulk $d$-wave superconductor, 
then one ought to have a
half-integral flux quantum trapped in this ring at low
temperatures.  More complicated scenarios such as those 
pictured in
Figs. 7a, 7b, and 7c can be investigated by varying the tricrystal junction
angles.  Note that for each case in these figures for which
$I_c(\phi_0^*)=0$, the relevant $I_c$ changes sign at $\phi_0^*$.
When only one junction is studied, the choice of
 phase factor across the junction is arbitrary, and that which
minimizes the free energy of the junction changes $I_c$ to $|I_c|$ when
$I_c<0$.  But, for a tricrystal (or tetracrystal), the
 relative phases across the junctions are fixed, due to the loop.

 The free energy of a ring with $n$ Josephson junctions is
given by
\begin{equation}
F=LI^2/2-(\Phi_0/2\pi)\sum_{i=1}^nI_{ci,i+1}
\cos(\Delta\varphi_{i,i+1}),
\end{equation}
where
\begin{eqnarray}
\sum_{i=1}^n\Delta\varphi_{i,i+1}&=&-2\pi\Phi/\Phi_0\cr
& &\cr
LI^2/2&=&(\Phi-\Phi_{\rm ext})^2/2L.
\end{eqnarray}
These were only written down for Josephson junctions
in the $ab$-plane,
as in the tricrystal experiments of Kirtley and
Tsuei. \cite{Kirtley,AGL,SR}  However, they also apply to the
case of a ring of $c$-axis
Josephson junctions.    One has only to be careful that the
self-inductance $L$ is	sufficiently large that
$\beta=LI_{c}/\Phi_0>>1$, where $I_{c}$ is the
minimum value of the critical currents
across the junctions. \cite{Tesche}
 For the $ab$-plane junctions, although $J_c$ is large, the junction area is small enough that $I_{c}\approx 
2$ mA might not be large enough relative to $L$. \cite{Kirtley}
 Although $J_c$ for the $c$-axis junctions is much smaller than for the $ab$-junctions, the much larger 
junction areas make $I_{c}\approx 20-200$ mA, typically larger 
than for
the $ab$-plane junctions. \cite{Li3}   Since  $L$  is proportional
to the area inside the ring, by making the ring out of large
single crystals, it should be possible to construct rings with
$\beta>>1$.  However, in order not to lose the SQUID sensitivity, 
 the central area inside the ring should be nearly as 
small as 
for the $ab$-plane tricrystal experiments, with a rather flat insulating support for the SQUID microscope.  

We remark that in the previous ($ab$-plane) tricrystal experiments, the dependence of the critical current $I_{c}^{ij}$ upon the misalignment angles $\theta_i$  and $\theta_j$ was assumed to be proportional to either $\cos(2\theta_i)\cos(2\theta_j)$ or $\cos[2(\theta_i+\theta_j)]$, for coherent or incoherent $ab$-plane tunneling, respectively.  However, all published experiments on YBCO showed  an {\it exponential} dependence upon the misalignment angle $\theta=\theta_i+\theta_j$. \cite{Dimos,Gross,Ivanov,Larbalestier}  In the range of misalignment angles used in those $ab$-plane tricrystal experiments, both of the above theoretical predictions are inaccurate by several orders of magnitude. \cite{Kirtley}

Gurevich and Pashitskii showed recently that the exponential dependence on misalignment angle  observed experimentally in such YBCO $ab$-plane  junctions is completely unrelated to any purported OP symmetry, as they were able to fit the data equally well with $s$-wave and $d_{x^2-y^2}$-wave OP models, assuming the presence of insulating dislocation cores and OP suppression at the grain boundaries.  \cite{GP}  Such grain boundary suppression is likely to be accompanied by oxygen stoichiometry variation and hence local magnetic moments, which could be responsible for the $\pi$-junctions apparently observed in those experiments. \cite{Larbalestier} Such problems are evidently not present in the $c$-axis twist experiments, however. \cite{Li3} Thus, our proposed new type of tricrystal (or tetracrystal) experiment is several orders of magnitude more reliable than is the $ab$-plane tricrystal experiment.

\section{conclusions}
We have presented tables of the allowed OP eigenfunctions for
tetragonal cuprates such as Hg1201 and NCCO, and
for the orthorhombic cuprates YBCO and BSCCO.  For
 tetragonal crystals, the possible spin singlet OPs are
$|s\rangle$, $|d_{x^2-y^2}\rangle$, $|d_{xy}\rangle$, and
 $|g_{xy(x^2-y^2)}\rangle$, the eigenfunctions of which are 
listed in Tables II and III.
For YBCO, the spin singlet OPs  are
$|s+d_{x^2-y^2}\rangle$ and $|d_{xy}+g_{xy(x^2-y^2)}\rangle$, with
eigenfunctions listed
in Table IV.
For BSCCO, the spin
singlet OPs are $|s+d_{xy}\rangle$ and 
$|d_{x^2-y^2}+g_{xy(x^2-y^2)}\rangle$, with eigenfunctions listed in Table V.

We distinguished between compatible OP components, which are
elements of the same OP basis set, and incompatible OP components
 which are elements of different OP basis sets.  Specifically, the
$s$-wave and $d_{x^2-y^2}$-wave OP components are
compatible in YBCO, but incompatible in BSCCO.
Compatible OP components have amplitudes which are in phase each other, and do
not require the occurrence of a second phase transition to
exist
below $T_c$.
Neither do any of them	require a second transition
 to become large at low $T$.
 Incompatible OP components, on the other hand, do not mix
just below $T_c$.  Whichever has the higher bare $T_c$ value
dominates just below $T_c$, and the other can only become
non-vanishing below a second phase transition.	These
incompatible
OP components are then $\pi/2$ out of phase with each
other below the second transition.

In addition,
we evaluated the  dependence of the $c$-axis critical current
upon the twist angle $\phi_0$ for each of the compatible OP
 eigenfunctions appropriate for each of these crystal
symmetries.
We have discussed these results with regard to the very
 recent results of Li {\it et al.} on $c$-axis twist junctions
 of BSCCO.
\cite{Li2,Li3}	These experiments do not show any $\phi_0$
dependence of the $c$-axis critical current density just below $T_c$.
 As a minimum,
these experiments indicate that the OP is 
$|s+d_{xy}\rangle$.

If the intertwist tunneling were predominantly coherent,
this
would further imply that the OP was very nearly isotropic,
having
 an $\ell=0$ $s$-wave form, with any additional components
of the  $|s+d_{xy}\rangle$  OP being too small  to
observe.  However, the more likely
scenario (for BSCCO, especially) is that the intertwist
tunneling is  entirely incoherent, and strongly dominated
by the	$s$-wave incoherent tunneling
 matrix element, such that $|J|^2=0$ and
$\tau_{\perp0}/\tau_{\perp\ell}\ll1$ for $\ell\ne0$.  In this
 case, $\tilde{\eta}_{\ell}\approx\delta_{\ell 0}$ in Table VII,
 and the $c$-axis twist experiments only prove
that the OP contains an $\ell=0$, $s$-wave component, and is
thus  $|s+d_{xy}\rangle$.	As such, the experiments  do
not rule
out the possibility of OP nodes.  Such a scenario {\it could} be
consistent with the quasiparticle gap observed in ARPES,
\cite{Ding,Shen}  with the  linear, low-$T$ in-plane magnetic penetration
depth measurements,
\cite{Sridhar} and with the Raman scattering.
\cite{Devereaux}

Assuming   the OP is  $|s+d_{xy}\rangle$,
if the ARPES
measurements
were indeed giving information related to the
superconducting
OP, indicating nodes  (or near-nodes)  at
 $\phi_{\bf k}=\pm45^{\circ}$, then the OP
would have to be similar to the ``extended $s$-wave''
form, $|\cos(2\phi_{\bf k})|$.	This would mean, of
course, that the sign of the OP was constant.  It would also
mean
that the amount of the  $d_{xy}$-wave OP
component relative to the magnitude of the entire OP
 would be
small.	However, if the OP did not actually change sign, then it
would be difficult to explain the impurity dependence of the
 in-plane penetration depth.  In addition, the linear, low-$T$
dependence of the out-of-plane penetration depth,
$\lambda_c(T)$ would be difficult
to explain if the intrinsic interlayer tunneling were incoherent.
\cite{Sridhar} Hence, some of these experiments
might require new theoretical explanations.

For either predominantly coherent of incoherent intertwist
tunneling, the
$c$-axis twist experiments of Li {\it et al.} still
demonstrate the {\it absence} of any purported
$d_{x^2-y^2}$-wave OP component near to $T_c$.	Since
$s$-wave and $d_{x^2-y^2}$ OP components are incompatible,
any $d_{x^2-y^2}$-wave component could {\it only} occur
below a second phase transition.  For its amplitude to be large
at low $T_c$, the magnitude of the specific heat anomaly
required at the second phase transition would have to be
comparable in magnitude to the anomaly at $T_c$, which
can be ruled out.
Thus, while these experiments do not preclude a small
$d_{x^2-y^2}$-wave component at low $T$, they are
inconsistent
with a small one near to $T_c$, and also with a large one
at low $T$.  As such,
these new phase-sensitive $c$-axis twist experiments
are  in direct conflict with
the recent tricrystal experiments of Tsuei and Kirtley
involving  BSCCO.
\cite{Kirtley2}

It remains to be seen if the same results can be
obtained in YBCO, NCCO, and Hg1201.  In YBCO,	since
the $s$-wave and $d_{x^2-y^2}$-wave OP components
compatible, one can not be as
clear about the possible mixing.  Nevertheless,
this experiment allows us to place limits upon the
purported mixing of $s$-wave and
$d_{x^2-y^2}$-wave OP components as a
function of $T$.  In no other phase-sensitive
experiment has this been yet possible in
the vicinity of $T_c$.

Finally, we  proposed that this technique can be
modified to prepare a tetracrystal, which can have the
desired characteristics of the tricrystal experiments of
Kirtley, Tsuei, and collaborators.  However, in this case,
the quality of the junctions would presumably be improved by
many orders of
magnitude. In addition, they do not require any expensive substrates,
so it should be relatively easy and inexpensive to reproduce
  these experiments  in other laboratories.

\acknowledgments

The authors  thank R. C. Dynes, K. Gray, R. Kleiner, Q. Li,  M. B. Maple, P. M{\"u}ller, and M. R. 
Norman for useful discussions.
This work was supported by 
USDOE-BES  Contract No. W-31-109-ENG-38,
by NATO  Collaborative Research Grant No. 960102,
and by the DFG through
the Graduiertenkolleg ``Physik nanostrukturierter
Festk\"orper.''

\begin{figure}[htb]
%\vspace*{-1.5cm}
%\epsfxsize=9cm
%\centerline{\epsffile{figure1.ps}}
%\vspace{-0.3cm}
\caption{Two-dimensional representations of a single CuO$_2$
plane. (a) Idealized positions in a tetragonal crystal of the Cu
(solid) and O (open circles) ion positions. (b)
 Group operations of a tetragonal crystal. (c) Group operations
 of YBCO. (d) Group operations of BSCCO}
\end{figure}

\begin{figure}[htb]
%\vspace*{-1.5cm}
%\epsfxsize=9cm
%\centerline{\epsffile{figure2.ps}}
%\vspace{-0.3cm}
\caption{$\ell\le4$ Even angular momentum OP basis functions
appropriate for Hg1201, BSCCO and YBCO.}
 \label{fig2}
\end{figure}

\begin{figure}[htb]
%\vspace*{-1.5cm}
%\epsfxsize=9cm
%\centerline{\epsffile{figure3.ps}}
%\vspace{-0.3cm}
\caption{Illustration of a $c$-axis twist junction with
 twist
angle $\phi_0$. }\label{fig3}
\end{figure}

\begin{figure}[htb]
%\vspace*{-1.5cm}
%\epsfxsize=9cm
%\centerline{\epsffile{figure4a.ps}}
%\centerline{\epsffile{figure4b.ps}}
%\vspace{-0.3cm}
\caption{(a) Tight-binding primary (thick solid curves)  and secondary
(thin solid and dashed) Fermi surfaces of
BSCCO.	The
high symmetry points $\Gamma$, $Y$, $X$, and
$\overline{M}$
in the first (approximately tetragonal) Brillouin zone (BZ) are
indicated, and the exact ($\Gamma-Y$) and approximate ($\Gamma-X$)
mirror-plane symmetry diagonals are
represented by the thin straight lines.   (b) Two first BZs as
in (a) rotated 45$^{\circ}$ with respect to each
other.}\label{fig4}
\end{figure}

\begin{figure}
%\vspace*{-1.5cm}
%\epsfxsize=9cm
%\centerline{\epsffile{figure5.ps}}
%\vspace{-0.3cm}
\caption{Plot of $I_c(\phi_0)/I_c(0)$ for the case considered in
Ref. (24) of a dominant $d_{x^2-y^2}$-wave and subdominant
$d_{xy}$-wave OP, the relative amounts varying with layer index away
from the twist junction.  The parameters for these curves are
$T_{cB}/T_{cA} = 0.2$, $T_{cB}^{<}/T_{cA} = 0.1304$,
$\epsilon/6\beta_A = 0.5$, $\delta/6\beta_A = 0.1$, and the curves for
$\eta=\eta'=0.1$ and $\eta=\eta'=0.001$ are indicated.  Curves for
$t= T/T_{cA} =0.99, 0.9, 0.5$ are presented.}\label{fig5}
\end{figure} 

\begin{figure}
%\vspace*{-1.5cm}
%\epsfxsize=9cm
%\centerline{\epsffile{figure6.ps}}
%\vspace{-0.3cm}
\caption{Plot of $J_c^J(\phi_0)/J_c^S$ versus $2\phi_0/\pi$
for the simplest cases of pure $s$-wave (s), either $d$-wave
(d),
either $p$-wave (p), or either $g$-wave (g) OP contributions.}
\label{fig6}  \end{figure}

\begin{figure}
%\vspace*{-1.5cm}
%\epsfxsize=9cm
%\centerline{\epsffile{figure7a.ps}}
%\centerline{\epsffile{figure7b.ps}}
%\centerline{\epsffile{figure7c.ps}}
%\vspace{-0.3cm}
\caption{Theoretical $J_c^J(\phi_0)/J_c^S$ 
 versus $2\phi_0/\pi$ for
mixed OP contributions.  (a)  $A_1$ state with
$[A+B\cos(2\phi_0)+C\cos(4\phi_0)]/(A+B+C)$,  and
$B/A$, $C/A$ ratios given. (b) Same as in (a),
but
with small A values. (c) $A_2$ state with
 $[B\cos(2\phi_0)+C\cos(4\phi_0)]/(B+C)$, and $B/C$ ratios
given.}
\label{fig7}
\end{figure}

\begin{figure}
%\vspace*{-1.5cm}
%\epsfxsize=9cm
%\centerline{\epsffile{figure8.ps}}
%\vspace{-0.3cm}
\caption{ Proposed configuration of a $c$-axis
version
of the tricrystal ring experiment.  Dark crystal:  bottom.
Light crystal:
top.  Intermediate shading:  equal thickness crystals.
Arrows indicate the direction of a given single crystal axis.}\label{fig8}
\end{figure}
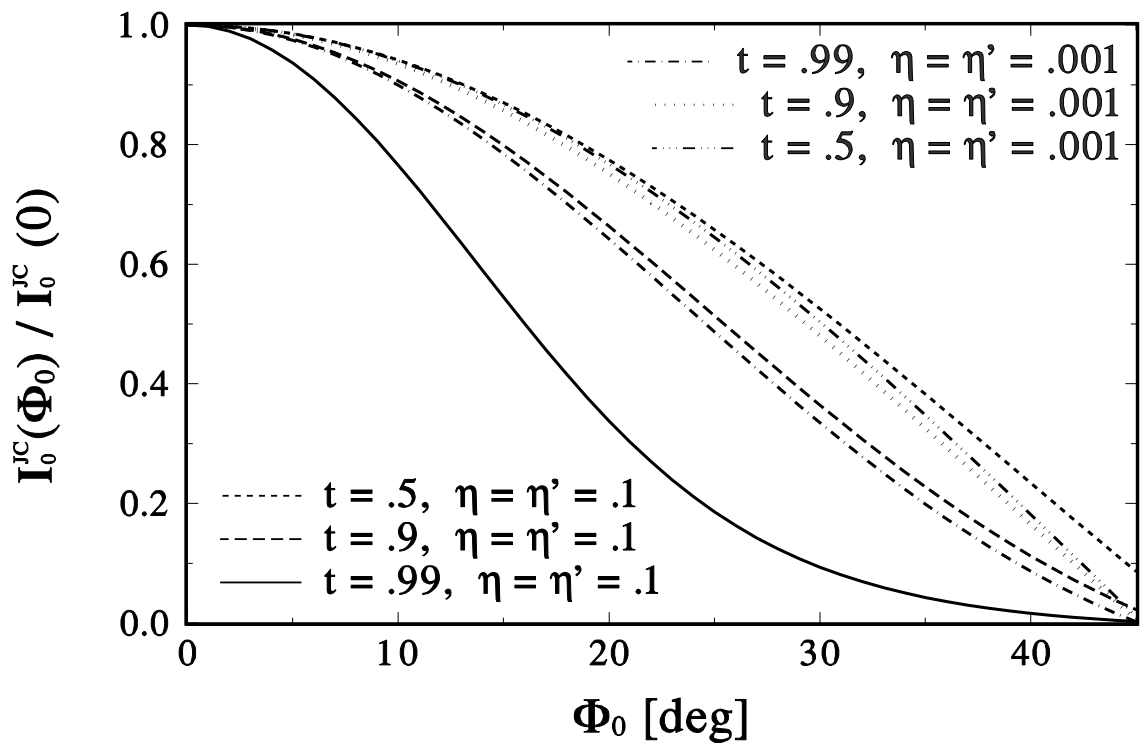
\end{document}